\documentclass[referee]{raa}
\usepackage{graphicx,times}
\usepackage{natbib}
\usepackage{amssymb,amsmath}
\usepackage{float}
\usepackage{subcaption}
\usepackage{graphics}
\usepackage{caption}
\usepackage{tabularx}
\usepackage{multirow}
\usepackage{booktabs}
\bibpunct{(}{)}{;}{a}{}{,}

\usepackage[pagebackref=true]{hyperref}

\begin{document}

   \title{Comprehensive study of the blazars from Fermi-LAT LCR: The log-normal flux distribution and linear RMS-Flux relation}

 \volnopage{ {\bf 20XX} Vol.\ {\bf X} No. {\bf XX}, 000--000}
   \setcounter{page}{1}

   \author{Na Wang \inst{1, 2, 3}, Ting-Feng Yi\inst{1, 2, 3}, Liang Wang\inst{1}, Li-Sheng Mao\inst{1}, Zhi-Yuan Pu
      \inst{1}, Gong-Ming Ning\inst{1}, Wei-Tian Huang\inst{1}, He Lu\inst{1}, Shun Zhang\inst{1}, Yu-Tong Chen\inst{1}, Liang Dong\inst{3}
   }

   \institute{ Department of Physics, Yunnan Normal University, Kunming 650500,
China; {\it yitingfeng@ynnu.edu.cn}\\
\and
Guangxi Key Laboratory for the Relativistic Astrophysics, Nanning 530004,  China \\
\and
Yunnan Province China-Malaysia HF-VHF Advanced Radio Astronomy Technology International Joint Laboratory, Kunming 650011, China\\
\vs \no
   {\small Received 20XX Month Day; accepted 20XX Month Day}
}

\abstract{Fermi-LAT LCR provide continuous and regularly-sampled gamma-ray light curves, spanning about 14 years, for a large sample of blazars.
The log-normal flux distribution and linear RMS-Flux relation of the light curves for a few of Fermi blazar have been examined in previous studies. However, the probability that blazars exhibit log-normal flux distribution and linear RMS-Flux relation in their gamma-ray light curves has not been systematically explored. In this study, we comprehensively research on the distribution of $\gamma$-ray flux and the statistical characteristics on a large sample of 1414 variable blazars from the Fermi-LAT LCR catalog, including 572 FSRQs, 477 BL Lacs, and 365 BCUs, and statistically compare their flux distributions with normal and log-normal distributions. The results indicate that the probability of not reject log-normal is $42.05\%$ for the large sample, and there is still $2.05\%$ probability of not reject normality, based on the joint of Kolmogorov-Smirnov, Shapiro-Wilk  and Normality tests. We further find that the probability that BL Lacs conforms to the log-normal distribution is higher than that of FSRQs.
Besides, after removing sources with less than 200 data points from this large sample, a sample of 549 blazars, which is still a large sample comparing to the previous studies, was obtained. Basing on dividing the light curves into segments every 20 points (or 40 points, or one year), we fitted the linear RMS-Flux relation of this three different sets,
and found that the Pearson correlation coefficients are all close to 1 of the most blazars. This result indicates a strong linear correlation between the RMS and the flux of this 549 blazars. The log-normal distribution and linear RMS-Flux relation indicate that the variability of $\gamma$-ray flux for most blazars is non-linear and multiplicative process.
}

   \authorrunning{Na Wang et al.}            
   \titlerunning{Comprehensive properties of the variable blazars}  
   \maketitle

%
\section{INTRODUCTION}           
\label{sect:intro}
Blazars, a special subclass of Active Galactic Nuclei (AGNs), exhibit unique properties that are notably characterized by their highly relativistic jets which are oriented in the direction of the line of sight of the observers (\citealt{Urry+Padovani+1995, Jovanovi'c+etal+2023}).
These objects exhibit extreme observational properties including high luminosity, high polarization, fast variability.
Blazars are divided into two sub-classes: BL Lac object (BL Lac) and Flat-Spectrum Radio Quasars (FSRQ). The spectrum of BL Lacs have only  faint emission lines or no emission lines,  but FSRQs has strong emission lines (\citealt{Urry+Padovani+1995, Abdo+etal+2010}). In the past thirty years, with the development of gamma ray astronomy, it has become possible to study the temporal and spectral behavior of blazars in detail.
The Fermi telescope provided a platform for the exploring high-energy $\gamma$-ray sources outside the Galaxy (\citealt{Atwood+etal+2009, Razzano+etal+2009}). The latest catalog update includes over 5000 $\gamma$-ray sources above 4$\sigma$ significance, 60$\%$ of which are blazars (\citealt{Abdollahi+etal+2020}). 

The flux distribution of blazars has been studied in many literatures to understand the nature of emission processes (\citealt{Urry+Padovani+1995, Biteau+Giebels+2012a, Bhatta+Dhital+2020, Khatoon+etal+2020}).
A number of blazars observed by Fermi-LAT show log-normal distribution in their long-term $\gamma$-ray light curves 
(\citealt{Scarpa+Falomo+1997, Uttley+etal+2005, Ackermann+etal+2015, Shah+etal+2018, Romoli+etal+2018, Bhatta+Dhital+2020, Zhang+etal+2022, Zhang+etal+2023}).
In recent years, the progress of astronomical observations provides important conditions for analyzing long-term flux distributions. 
The multiple investigations have yielded log-normal distributions of X-ray flux existing in AGNs, X-ray binaries
(\citealt{Lyubarskii+1997, Uttley+McHardy+2001, Quilligan+etal+2002a, Quilligan+etal+2002b, McHardy+2010}).
Subsequently, a log-normal distribution of flux was discerned in the $\gamma$-ray light curves of PKS 1510-089 and PKS 2155-304, with a linear correlation linking the Root Mean Square (RMS) and the average flux (\citealt{Kushwaha+etal+2016, H.E.S.S.Collaboration+etal+2017}).

\cite{Lyutyj+Oknyanskij+1987} find that the linear correlation between X- ray flux and its corresponding variation implies a log-normal distribution in Seyfert galaxy.
In X-ray band, many AGNs exhibit both log-normal distribution and the RMS-Flux linear connection (\citealt{McHardy+etal+2004, Kushwaha+etal+2017}).
The proportional linear correlation denoted that the absolute amplitude of RMS and mean flux were linked, thereby infered that sources with greater luminosity evince greater degrees of RMS fluctuation.
Furthermore, this linear relationship may possess a more intrinsic and  fundamental than the PSD (Power Spectral Density) (\citealt{McClintock+etal+2003, Gleissner+etal+2004, Pottschmidt+etal+2004}).
It has always been controversial that the flux variability of blazars origined from the jet itself or the disk that could modulate the jet emission.
The RMS-Flux linear connection can promote the study of potential physical process changes in blazars. Therefore, the linear RMS-Flux relation is of great significance for the time variable analysis of the blazars, and it is helpful for us to understand the origin of variability of blazars.

In $\gamma$-ray band, this RMS-Flux relation and log-normal distribution was researched for some individual sources, such as Mrk 421 , Mrk 501, 1ES 1011+496 (\citealt{Tluczykont+etal+2010, Chakraborty+etal+2015, Sinha+etal+2017}).
Small sample of log-normal distribution were also investigated (\citealt{Kushwaha+etal+2017, Shah+etal+2018, Bhatta+Dhital+2020}).
However, There is still a lack of systematic research of flux distribution and RMS-Flux relation of the $\gamma$-ray of blazars.
In order to further systematically study the RMS-Flux relation and the flux distribution of the blazars. We will perform comprehensive study on $\gamma$-ray flux of the Fermi LAT blazars, basing on a large sample. From the 1525 variable sources (variable index more than 21.67) of the Fermi-LAT LCR catalog (\citealt{Abdollahi+etal+2023}), we obtained the large sample including 1414 variable blazars. They have been continuously monitored at a cadence of three days, making them well suited for time series analysis. The data samples are described in Sect. 2. The flux distributions will be fitted by both the normal and log-normal functions (Sect. 3). In Sect.4, we analyze and fit the RMS-Flux relation on the light curve by using three different binnings, and compare results statistically between the three subtypes of sources, FSRQs, BL Lacs and blazar candidate of uncertain type (BCUs). The results and possible implications are discussed in Sect. 5.


\section{DATA SAMPLES}
\label{sect:Obs}
In 2008, the Fermi Gamma-ray Space Telescope was launched with the Large Area Telescope (LAT) as the principal tool in its scientific repertoire.
It usually runs in scan mode and covers the entire thing
$\gamma$-ray photon events in the sky range from 20 MeV to \textgreater{300}GeV every $\sim$3 hours with a field of view of $\sim$ 2.4 steradian. The telescope provides continuous and nearly uniform observations of persistent and transient $\gamma$-ray sources which the longest and most uniformly sampled $\gamma$-ray data (\citealt{Abdo+etal+2009}).
The light curves on time interval  of 30 days (monthly), 7 days (weekly), and 3 days for 1414 variable blazars sources (variable index more than 21.67) in the Fermi-LAT LCR (\citealt{Abdollahi+etal+2023}).
Here, the light curves on the cadence of 3 days with minimum detection significance TS\textgreater{4} (2$\sigma$) from 2008 to now are chosen providing more data points for studying the RMS-Flux relation and flux distribution.
We further investigated or removed any outliers 
(photon flux \textgreater{$5\times10^{-5}$}) before using the data for detail analysis.
These blazars that can be classified into FSRQs, BL Lacs, and BCUs, accounting for 40.453\%, 33.734\%, and 25.813\% of the samples respectively.
We use the light curves of the blazars of these three types to study the flux distribution.
After removing sources with less than 200 data points (to reduce the influence to RMS-Flux from the small number of data points in per bin) from this large sample , we obtain a subsample including 41 BCUs, 236 BL Lacs, and 272 FSRQs for fitting RMS-Flux relation.

\section{FLUX DISTRIBUTION ANALYSIS}
\label{sect:Obs}
\begin{figure}[H]
  \centering
  \begin{minipage}{0.32\textwidth}
    \centering
    \includegraphics[width=\linewidth]{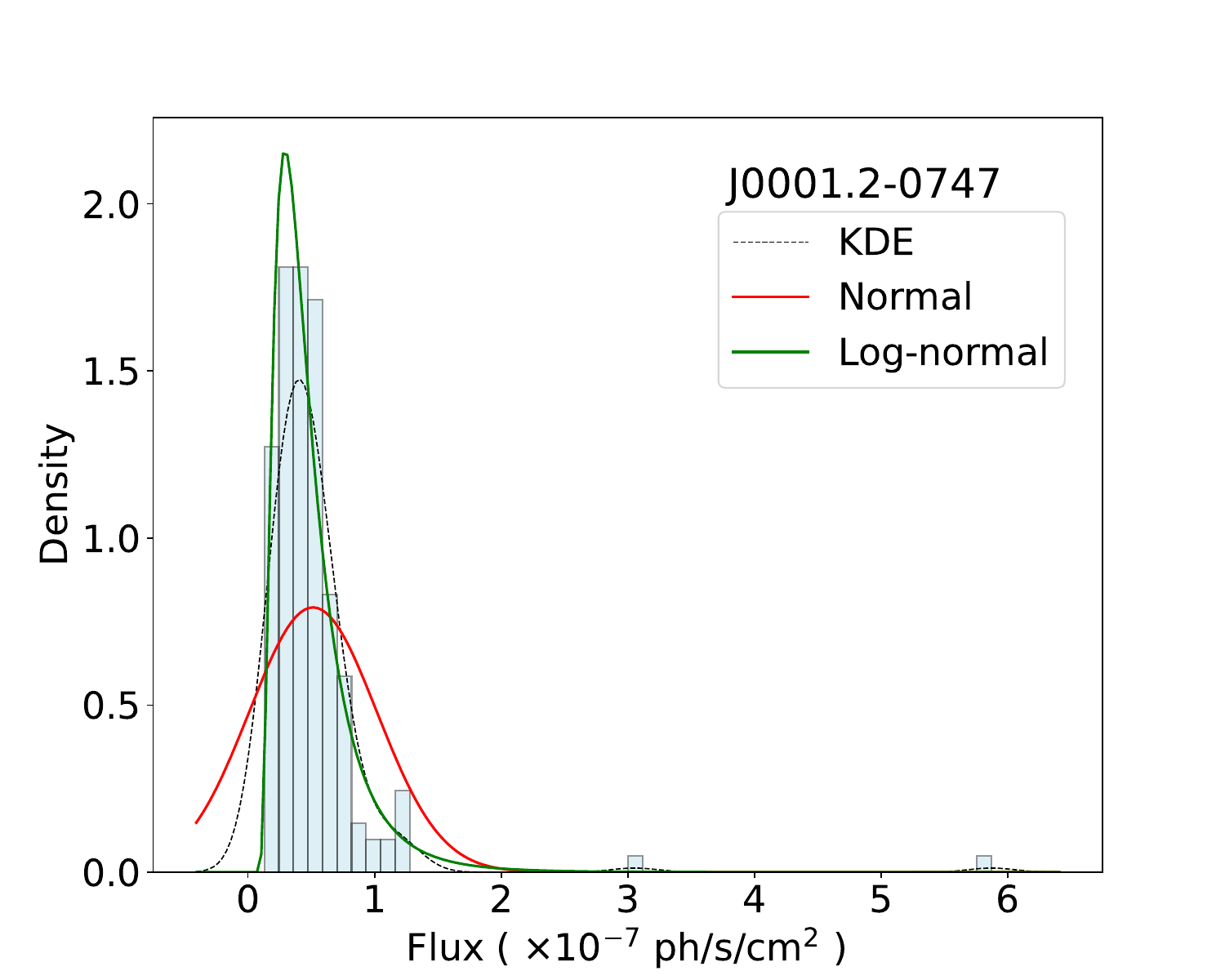}
  \end{minipage}
  \hfill
  \begin{minipage}{0.32\textwidth}
    \centering
    \includegraphics[width=\linewidth]{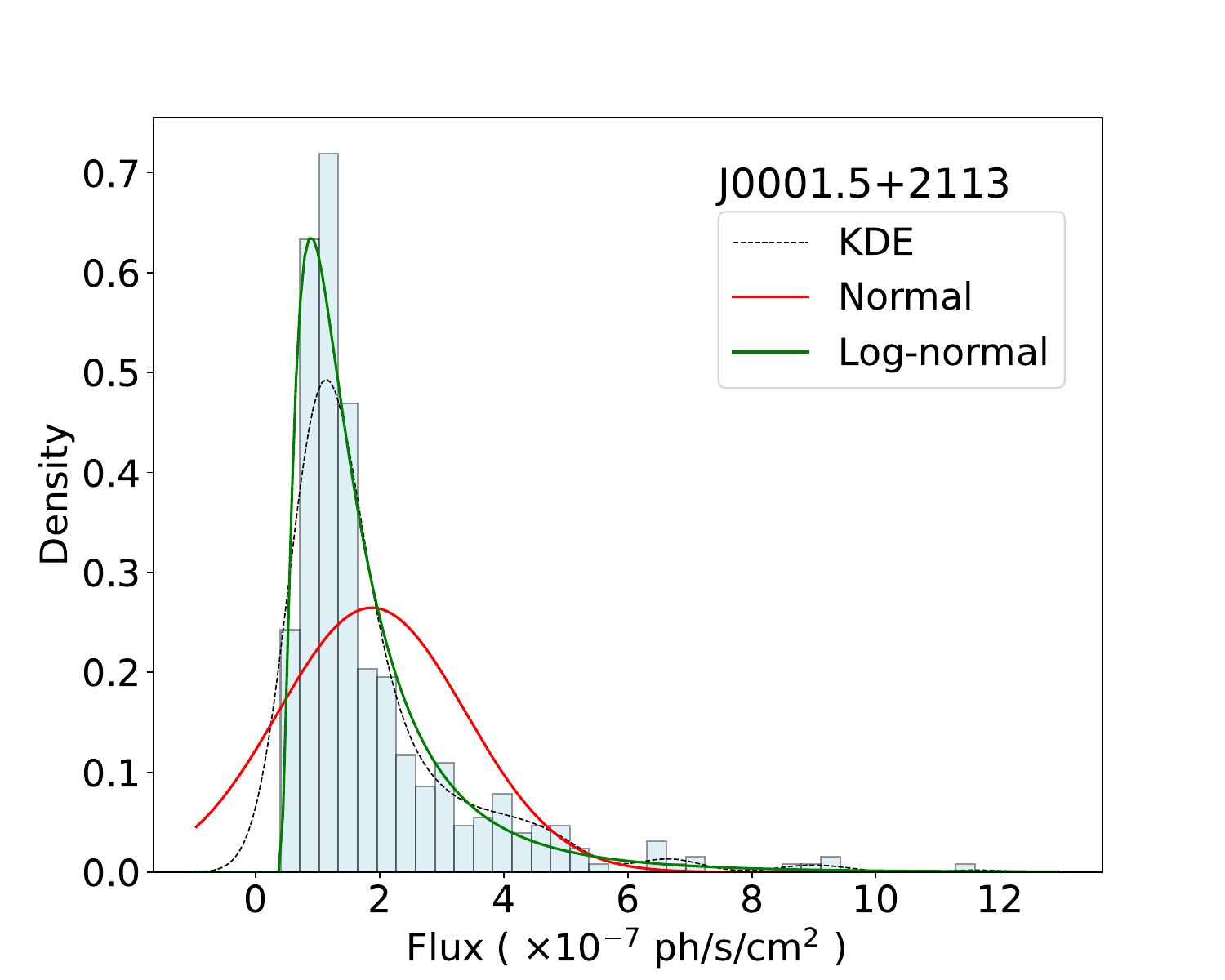}
  \end{minipage}
  \hfill
  \begin{minipage}{0.32\textwidth}
    \centering
    \includegraphics[width=\linewidth]{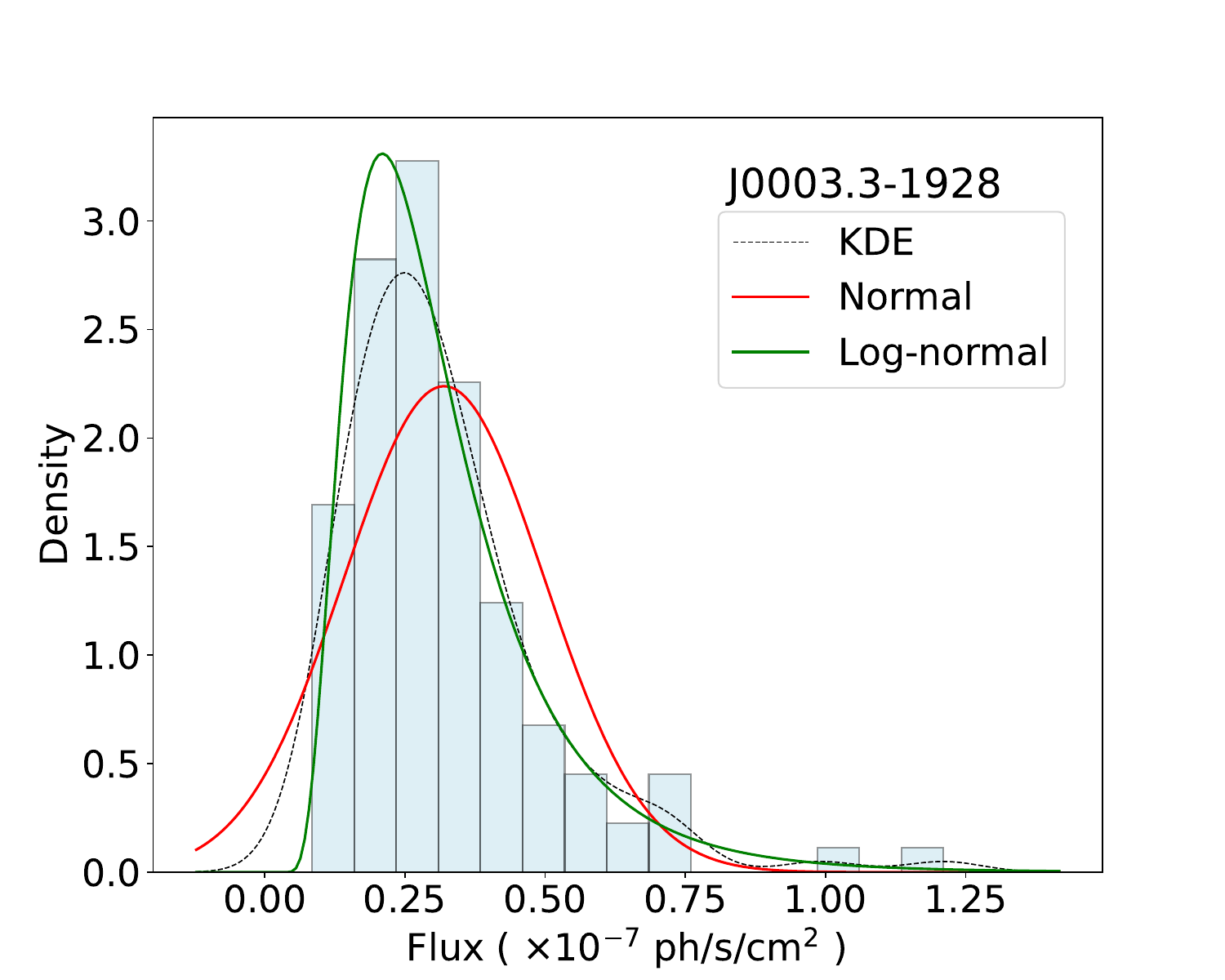}
  \end{minipage}
  \begin{minipage}{0.32\textwidth}
    \centering
    \includegraphics[width=\linewidth]{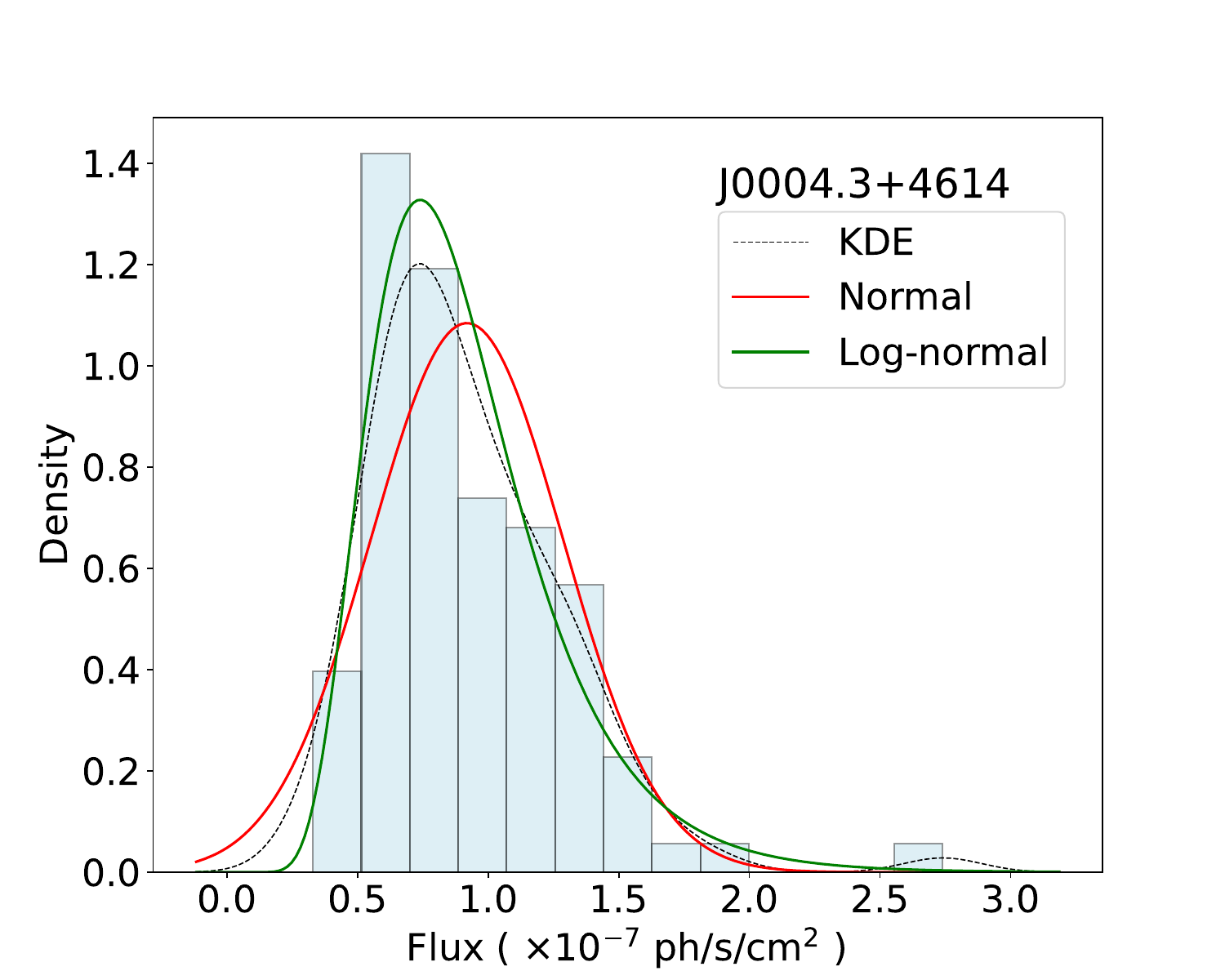}
  \end{minipage}
  \hfill
  \begin{minipage}{0.32\textwidth}
    \centering
    \includegraphics[width=\linewidth]{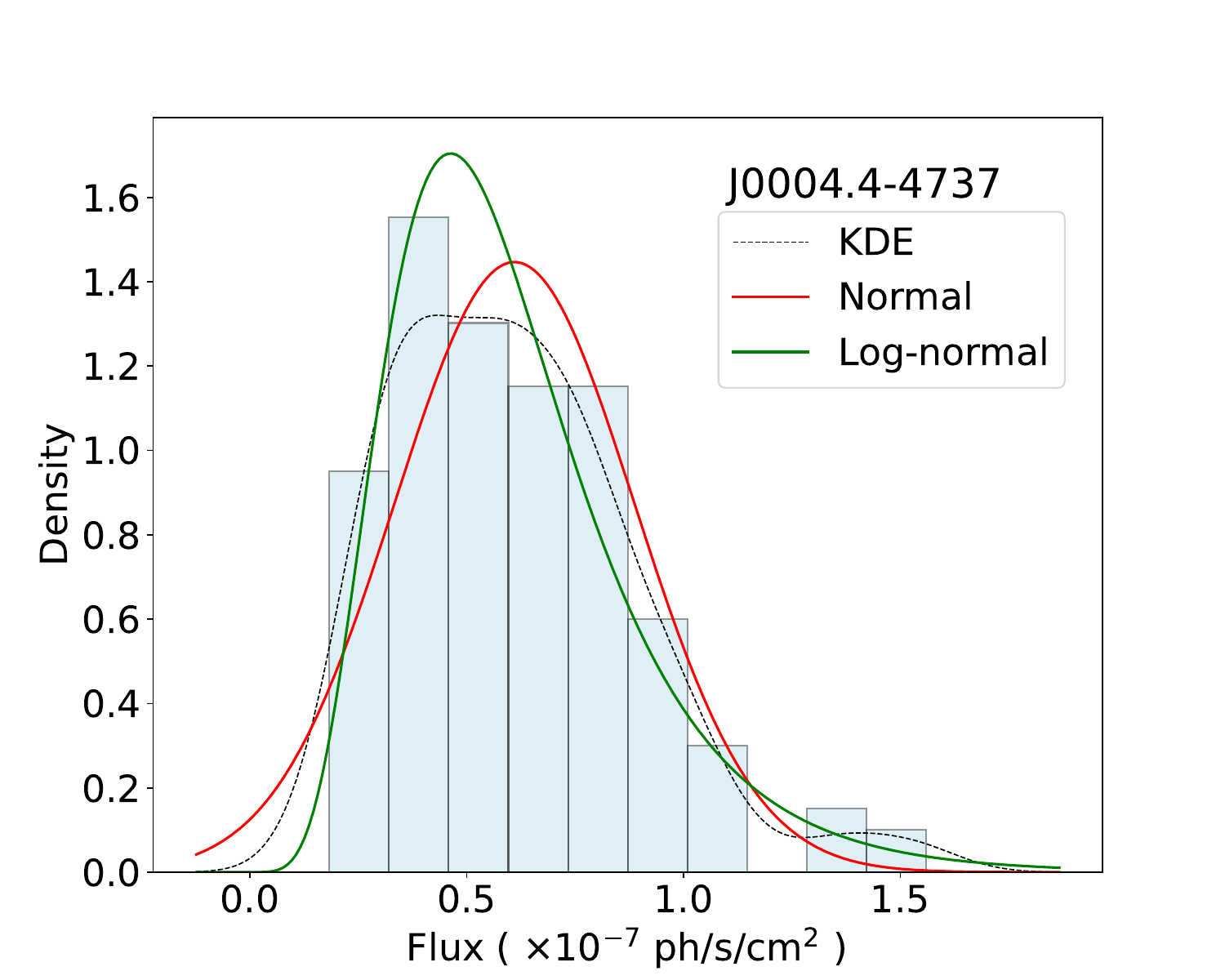}
  \end{minipage}
  \hfill
  \begin{minipage}{0.32\textwidth}
    \centering
    \includegraphics[width=\linewidth]{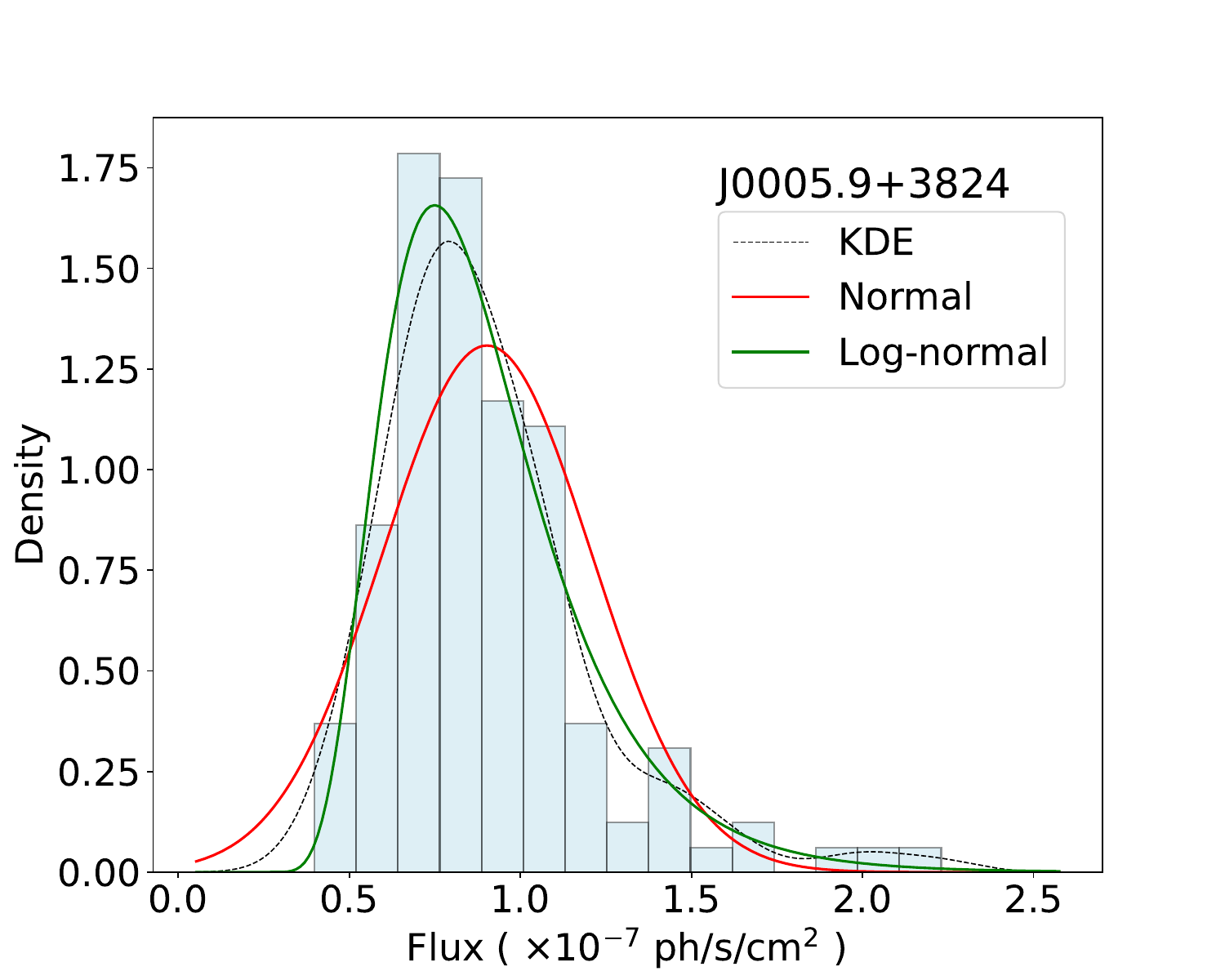}
  \end{minipage}
  \begin{minipage}{0.32\textwidth}
    \centering
    \includegraphics[width=\linewidth]{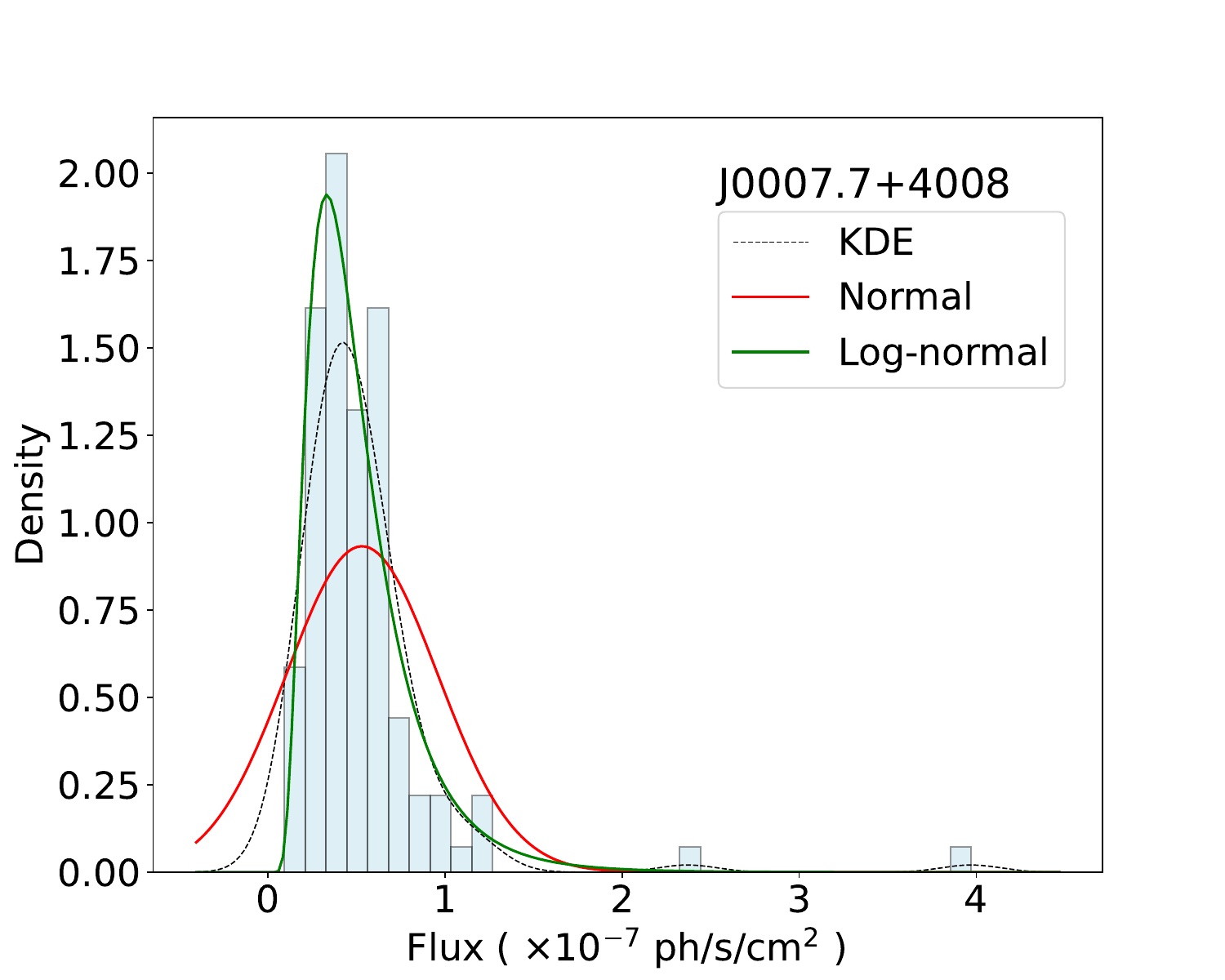}
  \end{minipage}
  \hfill
  \begin{minipage}{0.32\textwidth}
    \centering
    \includegraphics[width=\linewidth]{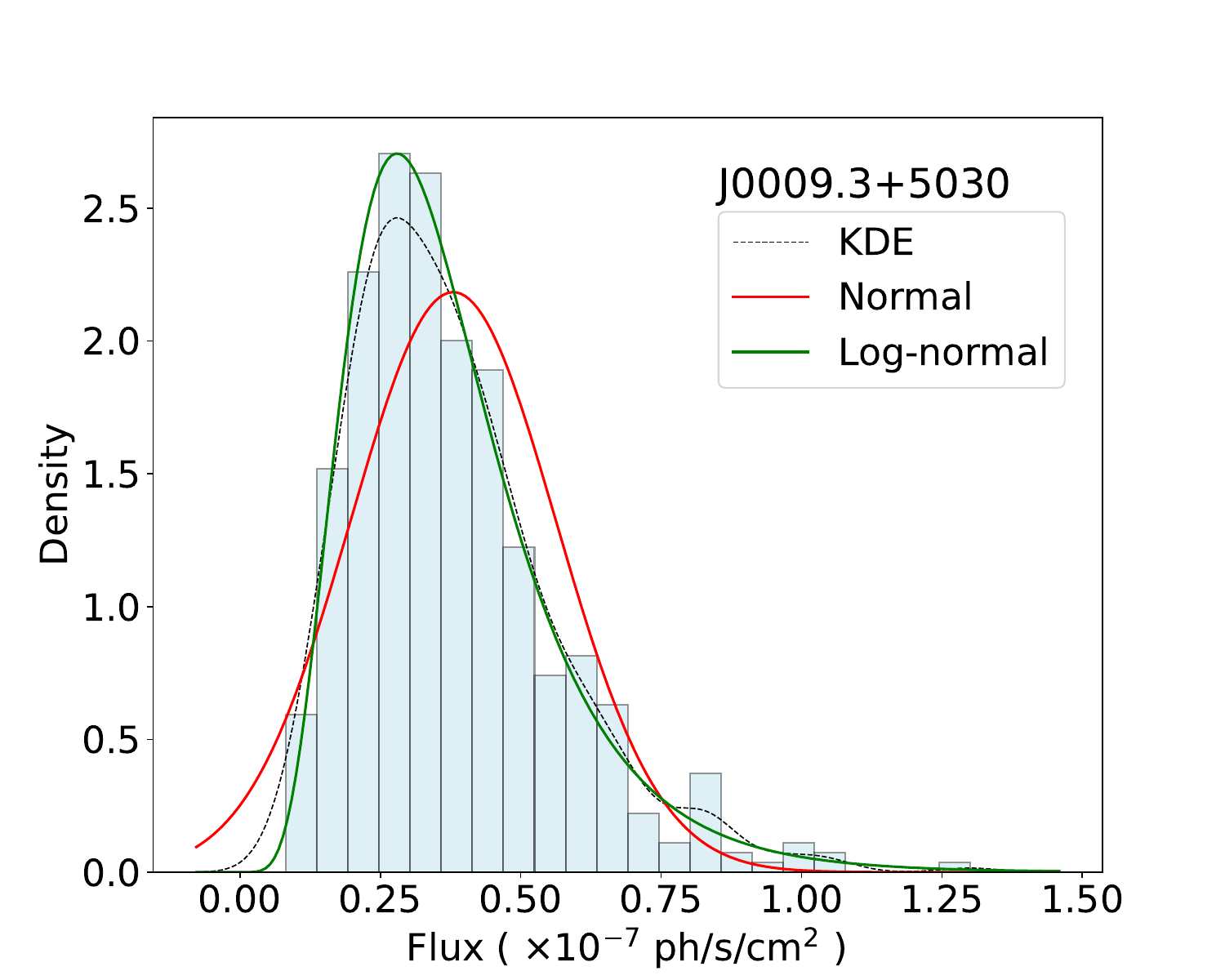}
  \end{minipage}
  \hfill
  \begin{minipage}{0.32\textwidth}
    \centering
    \includegraphics[width=\linewidth]{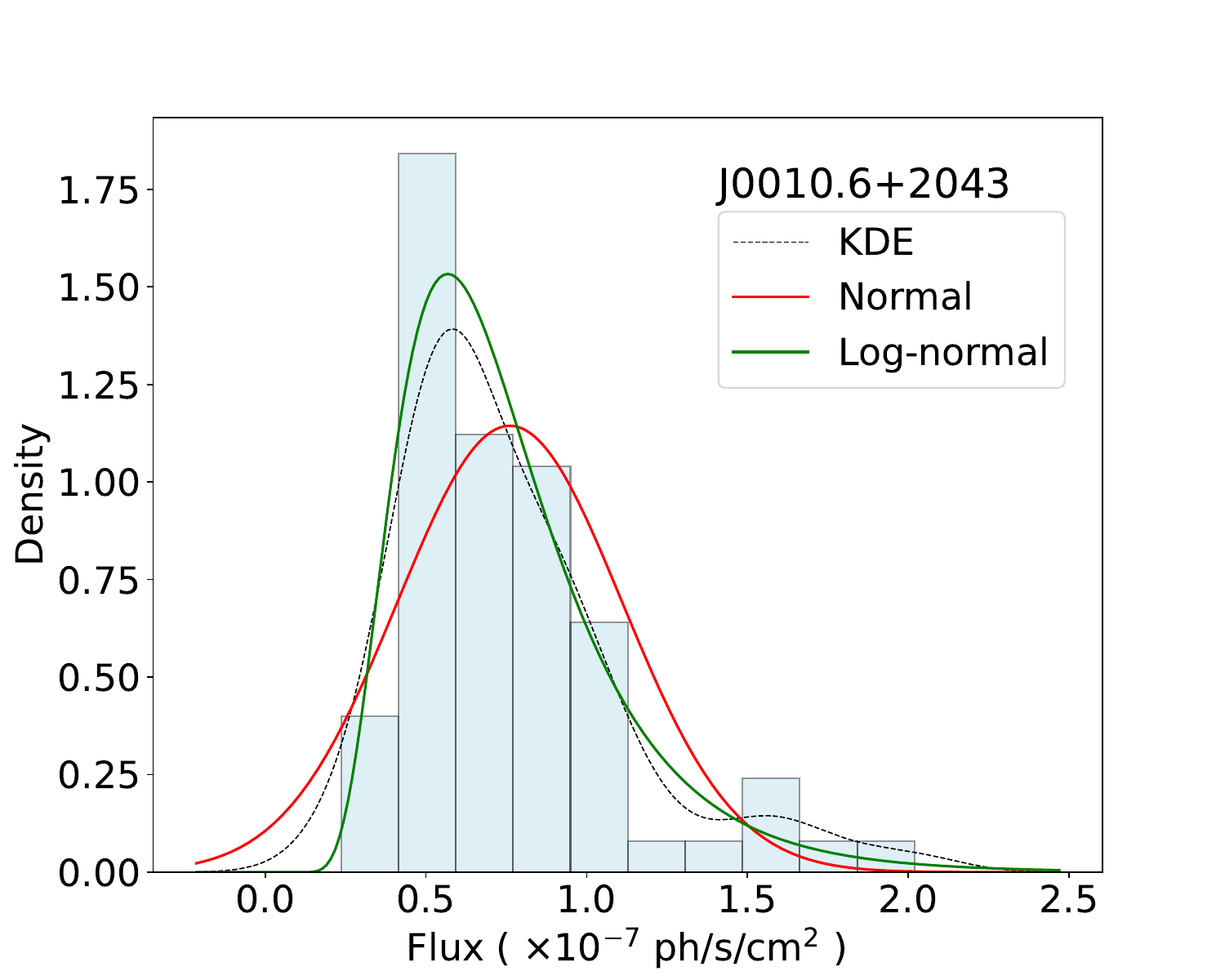}
  \end{minipage}
  \caption{The normal (red solid line) and log-normal (green solid line) function fitting to the  $\gamma$-ray flux distribution (hatched blue histogram) of the blazars, the black dotted line represents the kernel density estimation (KDE) curve. Here, the figure lists only the first 9 sources.  The entire figure is available in its entirety in machine-readable form.}
\label{Fig1}
\end{figure}

Histograms are a valuable tool for characterizing the flux distribution of blazar.
The flux histograms of 365 BCUs, 477 BL Lacs, and 572 FSRQs are shown in Figure~\ref{Fig1}.
We find that most histograms have a prominent peak and a high flux tail. 
To further understand the distributions, we fit them with normal and log-normal functions, and compared the results using reduced $\chi^2$.
The fit parameters together with the reduced $\chi^2$
for both distributions are shown in Table~\ref{tab1}.
We also compare the results by three tests, including the Kolmogorov-Smirnov (K-S), the Shapiro-Wilk (S-W) and Normality tests in Table~\ref{tab1}. We conduct statistical analysis on the results of three tests and reduced $\chi^2$ for BCUs, BL Lacs, and FSRQs.

We count the proportion of FSRQs, BL Lacs, and BCUs in Table~\ref{tab1} that did not reject the null hypothesis under the three tests of K-S, S-W, and Normality test, i.e., the proportion of p-values.
As shown in Table~\ref{tab2}, statistical analysis results of three tests are computed about 572 FSRQs to verify the normality or lognormality of the probability density function (PDF) from these histograms.
Out of the three tests, the percentage of do not reject normal distribution is 20.98\%, 1.92\%, and 2.45\%, respectively (p\textgreater{0.05}). On the other hand, the percentage of FSRQs do not reject the log-normal distribution of flux is 77.45\%, 37.41\%, and 37.41\%, respectively. Among the 477 selected BL Lacs, as shown in Table~\ref{tab2}, we perform a similar analysis process as for FSRQs. The precentage of BL Lacs dot not reject normal distribution is 21.59\%, 0.84\%, and 1.89\%, while 95.18\%, 52.41\%, and 52.83\% of BL Lacs do not reject log-normal of the flux distribution, respectively.

Apart from FSRQs and BL Lacs, there are 365 BCUs which lack reliable classification. To study the flux distribution characteristics of these 365 BCUs, we also analyze them as described above and the results of analysis show in Table~\ref{tab2}.
For the three different tests, we find that there are differences in the results of the three tests, among which the results of the K-S test are significantly different from the results of the S-W test and the Normality test, and the results of the S-W test and the Normality test are basically the same. 
In order to unify the results, we define the flux distritution of  blazars that do not reject the null hypothesis under all three tests as distributions that consist with the null hypothesis. 
Under this standard, out of 1414 variable blazars, flux distribution for 199 FSRQs, 230 BL Lacs, and 179 BCUs consist with the log-normal distribution. However, of these 1414 variable blazars, the flux distribution of 9 FSRQs, 4 BL Lacs, and 16 BCUs respectively conforms to the normal distribution.

\begin{table}[H]
\fontsize{7}{8}\selectfont
\caption{Parameters of log-normal and normal distribution fitting for the $\gamma$-ray flux distribution of the Fermi LAT Sources. The table lists only the first 9 blazars. This entire table is available in its machine-readable form.}\label{tab1}
\setlength{\tabcolsep}{0.035cm}
\begin{tabular}{cccccccccccccccc}
\hline\hline
\multirow{2}*{Name(type)} &\multicolumn{6}{c}{{Log-normal Fit}}& {} &\multicolumn{6}{c}{{Normal Fit}} \\
\cline{2-7}\cline{9-14}
 & Mean& $\sigma$ & K-S test(prob) & S-W test(prob) & Normlity(prob) & Reduced $\chi^2$ &  & Mean & $\sigma$ & K-S test(prob) & S-W test(prob) & Normlity(prob) & Reduced $\chi^2$ \\
\hline\hline
J0001.2-0747(bll) & -0.85 & 0.55 & $4.78\times10^{-1}$ & $7.44\times10^{-5}$ & $3.34\times10^{-7}$ & 35.53 & & 0.51 & 0.50 &$8.79\times10^{-9}$&$2.11\times10^{-23}$&$1.23\times10^{-63}$& \textgreater{}100                \\
J0001.5+2113(fsrq) & 0.41  & 0.62 & $1.64\times10^{-3}$ & $4.35\times10^{-9}$ & $2.70\times10^{-7}$ & 6.92 & & 1.87 & 1.51 &$7.78\times10^{-16}$&$1.45\times10^{-25}$&$1.00\times10^{-54}$& \textgreater{}100                \\
J0003.3-1928(fsrq) & -1.27 & 0.50 & $9.21\times10^{-1}$ & $9.65\times10^{-1}$ & $7.49\times10^{-1}$ & 0.79  & & 0.32 & 0.18 &$7.99\times10^{-3}$&$1.36\times10^{-9}$&$3.67\times10^{-14}$& \textgreater{}100                \\
J0004.3+4614(fsrq) & -0.16 & 0.38 & $9.41\times10^{-1}$ & $6.15\times10^{-1}$ & $6.82\times10^{-1}$ & 0.87  & & 0.92 & 0.37 &$2.53\times10^{-1}$&$2.03\times10^{-6}$&$3.76\times10^{-10}$& \textgreater{}100                \\
J0004.4-4737(fsrq) & -0.60 & 0.46 & $5.30\times10^{-1}$ & $1.50\times10^{-1}$ & $2.28\times10^{-1}$ & 1.42  & & 0.61 & 0.28 &$4.26\times10^{-1}$ & $1.35\times10^{-5}$&$2.80\times10^{-5}$& 4.90              \\
J0005.9+3824(bcu) & -0.15 & 0.31 & $8.42\times10^{-1}$ & $1.81\times10^{-1}$ & $5.71\times10^{-2}$ & 1.96  & & 0.90 & 0.31 &$5.14\times10^{-2}$ & $1.09\times10^{-8}$&$4.60\times10^{-12}$& 44.77             \\
J0007.7+4008(bll) & -0.79 & 0.54 & $3.66\times10^{-1}$ & $7.70\times10^{-3}$ & $7.43\times10^{-4}$ & 3.19  & &  0.53 & 0.43 &$9.53\times10^{-6}$ & $4.70\times10^{-17}$&$4.99\times10^{-36}$& \textgreater{}100            \\
J0009.3+5030(bll) & -1.08 & 0.48 & $8.28\times10^{-1}$ & $5.06\times10^{-1}$ & $3.51\times10^{-1}$ & 0.74  & &  0.38 & 0.18 &$7.46\times10^{-4}$ & $7.65\times10^{-15}$&$2.14\times10^{-23}$& 
50.41\\
J0010.6+2043(bll) & -0.36 & 0.42 & $3.80\times10^{-1}$ & $2.50\times10^{-1}$ & $4.95\times10^{-1}$ & 1.47  & &  0.76 & 0.35 &$4.01\times10^{-2}$ & $2.70\times10^{-6}$&$1.89\times10^{-6}$& 5.41           \\
\hline
\end{tabular}\\
\end{table}

\begin{table}[H]
\centering
\begin{minipage}[]{\textwidth}
\caption[]{The statistical results of normal and log-normal flux distribution of three subtypes of blazars.\label{tab2}}\end{minipage}
\setlength{\tabcolsep}{1pt}
\small
\begin{tabularx}{\textwidth}{c@{\hspace{2em}}c@{\hspace{2em}}c}
  \hline\noalign{\smallskip}
\multicolumn{3}{c}{FSRQs}                                         \\ \hline
Type of test& fraction of not reject normal (p\textgreater{0.05}) & fraction of not reject log-normal (p\textgreater{0.05}) \\\hline
K-S test      & 20.98\%      & 77.45\%      \\
S-W test      & 1.92\%       & 37.41\%      \\
Normality test   & 2.45\%       & 37.41\%       \\
joint of the three tests & 1.57\%   & 34.79\%         \\
 \hline\noalign{\smallskip}
\multicolumn{3}{c}{BL Lacs}              \\
 \hline\noalign{\smallskip}
K-S test      & 21.59\%      & 95.18\%      \\
S-W test      & 0.84\%       & 52.41\%     \\
Normality test   & 1.89\%       & 52.83\%       \\
joint of the three tests & 0.84\%    &  48.22\%        \\
 \hline\noalign{\smallskip}
\multicolumn{3}{c}{BCUs}                   \\
 \hline\noalign{\smallskip}
K-S test      & 43.01\%      & 92.60\%      \\
S-W test      & 4.93\%       & 53.97\%     \\
Normality test   & 6.03\%       & 53.15\%        \\
joint of the three tests &  4.38\%    &  49.04\%                 \\\hline
\end{tabularx}
\end{table}

Furthermore, based on the K-S, S-W, and Normality test results, the distribution of p-values was obtained. As shown in Figure2, the p-value distributions of the FSRQs, BL Lacs and BCUs are similar. These distributions reveal that the p-values of the K-S (Log-normal), S-W (Log-normal) and Normality (Log-normal) tests are predominantly concentrate in the range where p \textgreater{0.05}. Conversely, p-values of the three tests (Normal) are mainly concentrated in the range where p \textless{0.05}. Generally, the more the flux distribution of blazars in the sample fits the log-normal distribution, the less flux distribution of blazars fits the normal distribution. These consistent results indicate that the flux distributions of the three subtypes of blazars prefer log-normal distribution over normal distribution.

\begin{figure}[H]
  \centering
  \begin{minipage}{0.32\textwidth}
    \centering
    \includegraphics[width=5.7cm]{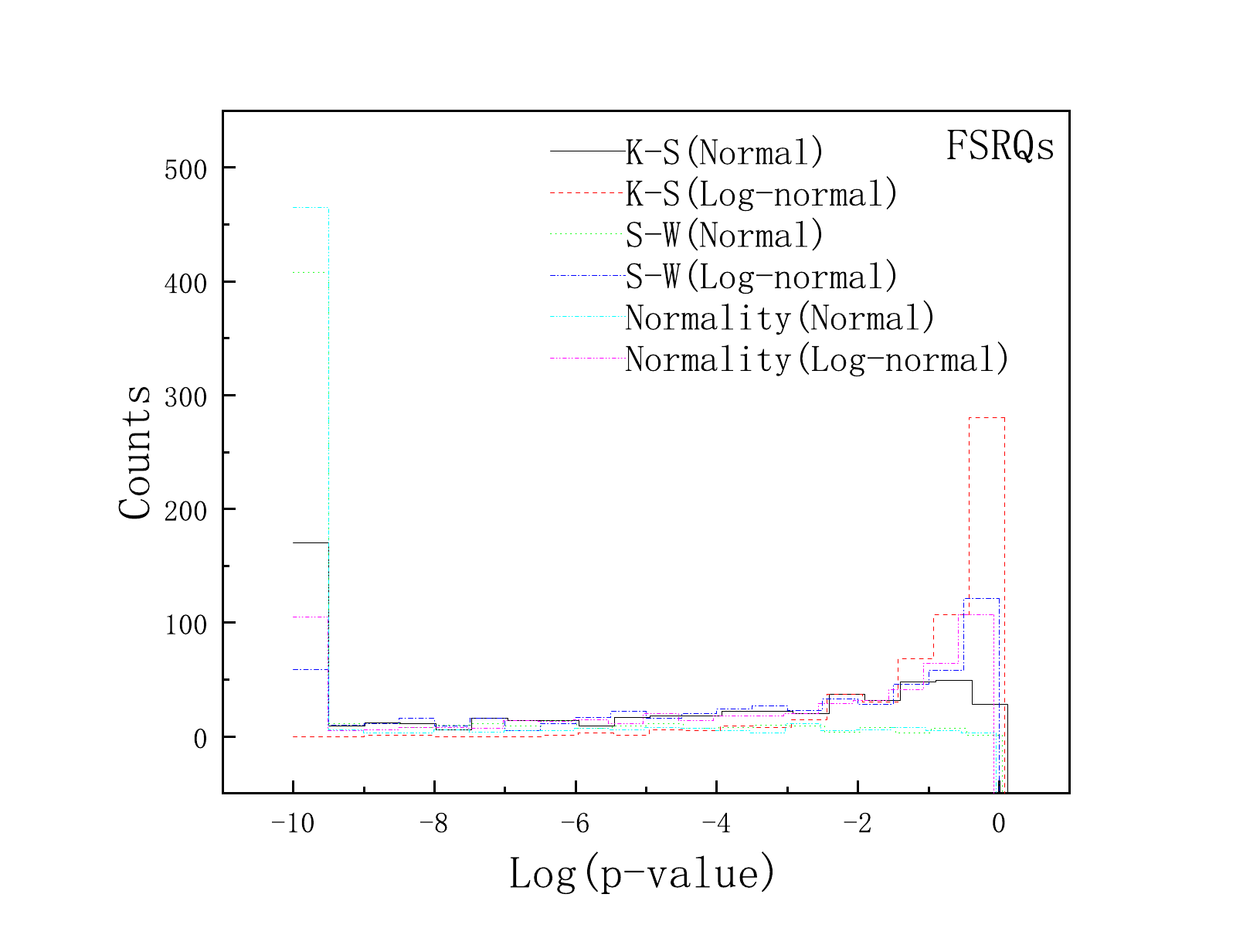}
    \label{FSRQs}
  \end{minipage}
  \hfill
  \begin{minipage}{0.32\textwidth}
    \centering
    \includegraphics[width=5.7cm]{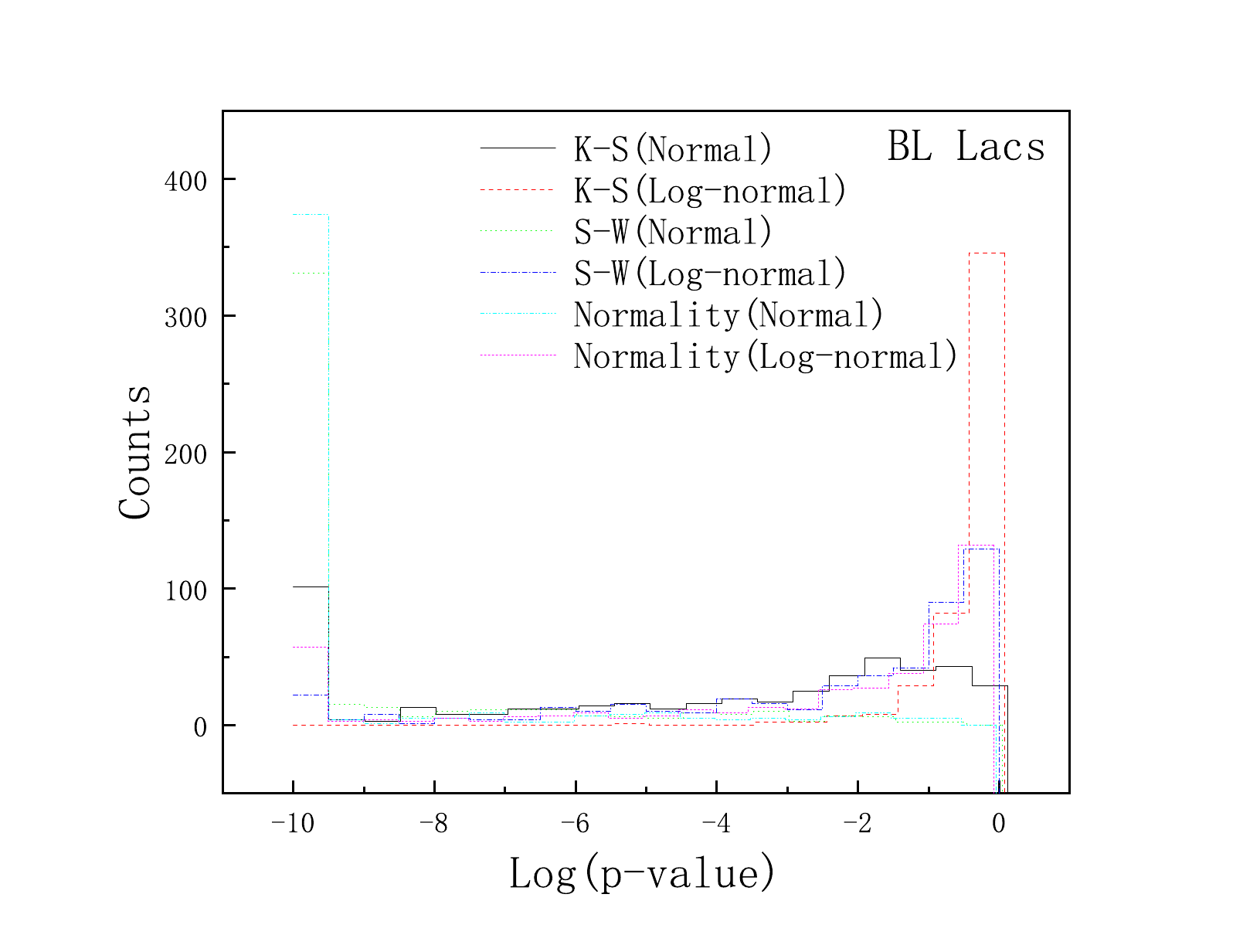}
    \label{FSRQs}
  \end{minipage}
  \hfill
  \begin{minipage}{0.32\textwidth}
    \centering
    \includegraphics[width=5.7cm]{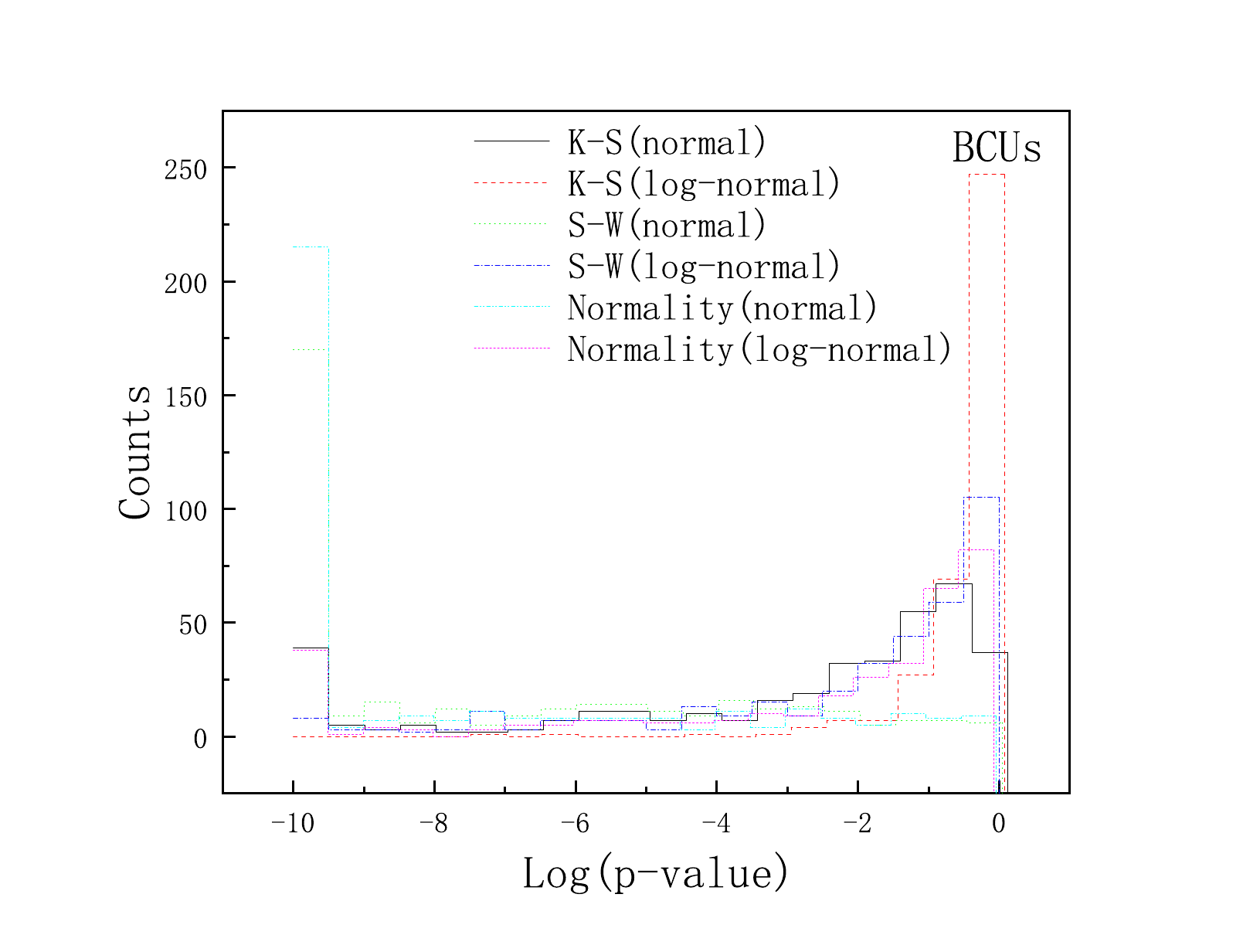}
    \label{FSRQs}
  \end{minipage}
  \caption{Distributions of p-value of K-S, S-W, and Normality tests for lognormality and normality of the flux distribution from blazars, respectively. The left, center, and right panels represent the statistical results for FSRQs, BL Lacs, and BCUs, respectively.}
\label{Fig2}
\end{figure}

The results in Table~\ref{tab2} and Figure~\ref{Fig2} of the three tests indicate that most of the blazars prefer to log-normal distribution.
According to the comparison of flux distribution from three test results, the probability that BL Lacs consist with the log-normal distribution is higher than that of FSRQs, while the probability that BCUs conform to the log-normal distribution is similar to that of BL Lacs.
The histograms and fits are shown in Figure~\ref{Fig1}.

In the multiplicative model, the flux naturally generates a log-normal distribution.
If we let $\phi$ be the product of flux A in a large number of isotropic regions that is,
\begin{equation}\label{eq1}
  \phi=\prod_{i=1}^{n}(A_n)
\end{equation}
\begin{equation}\label{eq2}
  log(\phi)=log(\prod_{i=1}^{n}(A_n))
\end{equation}
\begin{equation}\label{eq3}
  log(\phi)=\sum_{i=1}^{n}log(A_n)
\end{equation}
Then the log$\phi$ is the sum of logA. Pursuant to the Central Limit Theorem (CLT), it can be adduced that the logarithmic value of $\phi$ ensues a normal distribution, or in other words, $\phi$ adheres to a distribution which is log-normal (\citealt{Biteau+Giebels+2012b}).
Assuming in the context of a log-normal distribution, the observed flux $\phi$ can be expressed as a function of a underlying variable x, that is, $\phi$ = f (x). In this distribution, the function $f$ follows an exponential pattern while the variable $x$ conforms to a normal distribution. Fluctuations in $x$ in the form of a slight deviation from a given value, namely $\delta$x, result in corresponding deviations in $\phi$ around $f(x_0)$, $\delta$$\phi$  (\citealt{Biteau+Giebels+2012a}). It is noteworthy that the flux's variance is inexorably dependent on that of the variable $x$, and the variance of the flux is
\begin{equation}\label{eq4}
  \delta\phi^2=[\frac{\partial f}{\partial x}(x_0)]^2\delta x^2.
\end{equation}
When the flux deviation is proportional to f($x_0$),
the relation can be equivalent to
\begin{equation}\label{eq5}
  f(x_o)^2\propto[\frac{\partial f}{\partial x}(x_0)]^2\delta x^2.
\end{equation}
This equation is one of the definitions of the exponential function. The flux is proportional to its RMS only when the flux is an exponent of the underlying variable.
Therefore, the RMS-Flux relation can be interpreted as a consequence of a log-normal distribution of the flux (\citealt{Uttley+etal+2005,Biteau+Giebels+2012a,Biteau+Giebels+2012b}).

\section{RMS-Flux relation}
\begin{figure}[H]
  \centering
  \begin{minipage}{0.32\textwidth}
    \centering
    \includegraphics[width=\linewidth]{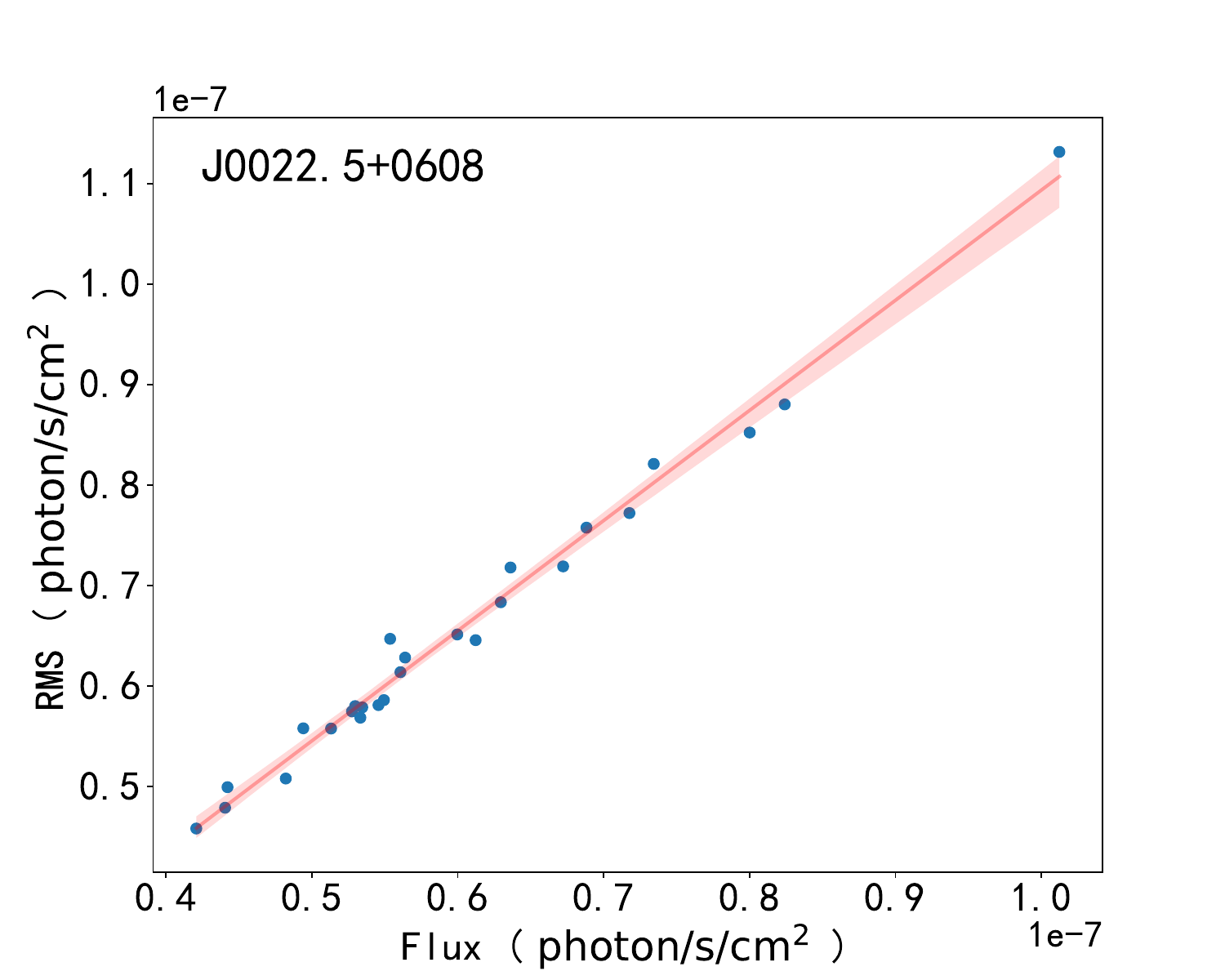}
  \end{minipage}
  \hfill
  \begin{minipage}{0.32\textwidth}
    \centering
    \includegraphics[width=\linewidth]{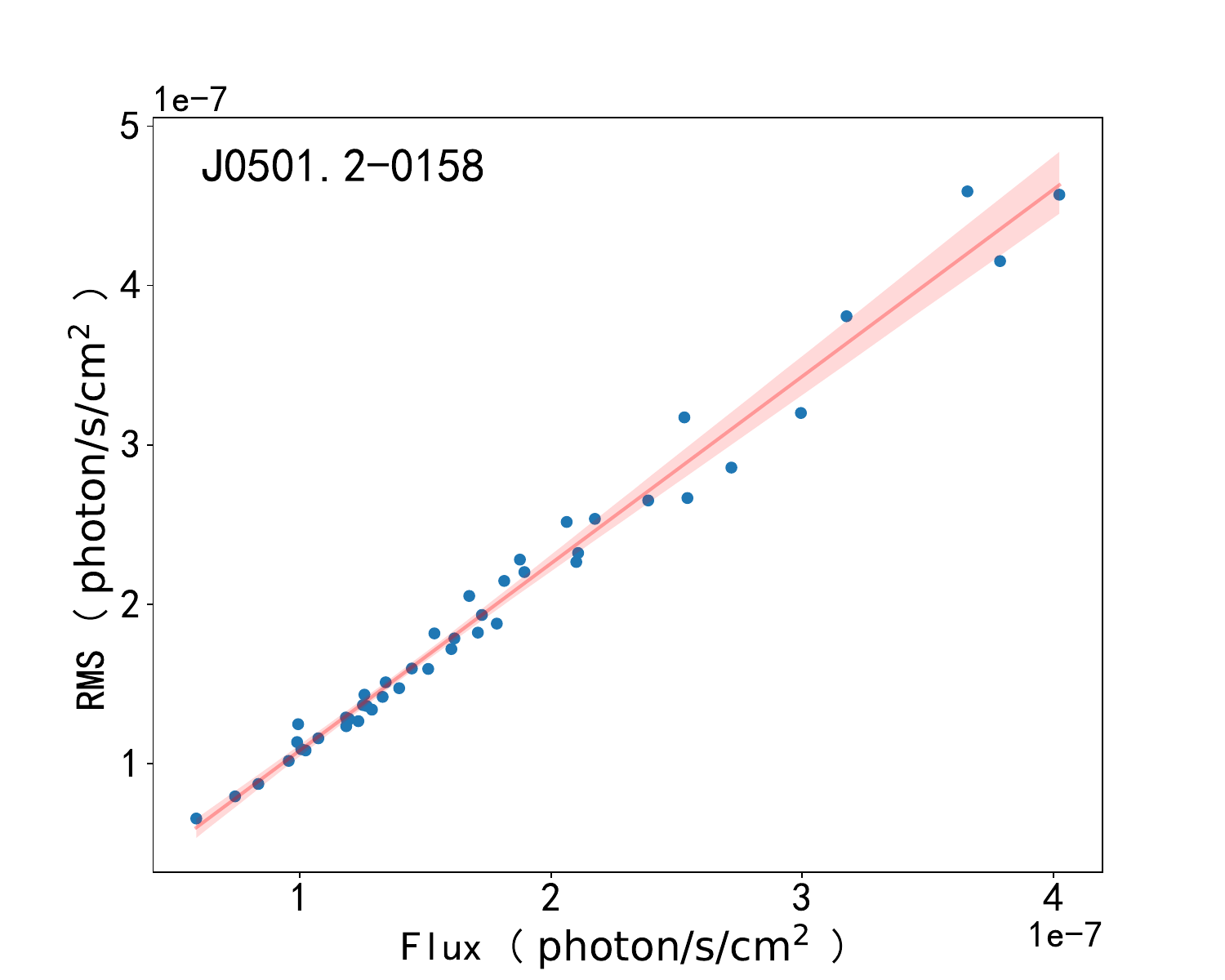}
  \end{minipage}
  \hfill
  \begin{minipage}{0.32\textwidth}
    \centering
    \includegraphics[width=\linewidth]{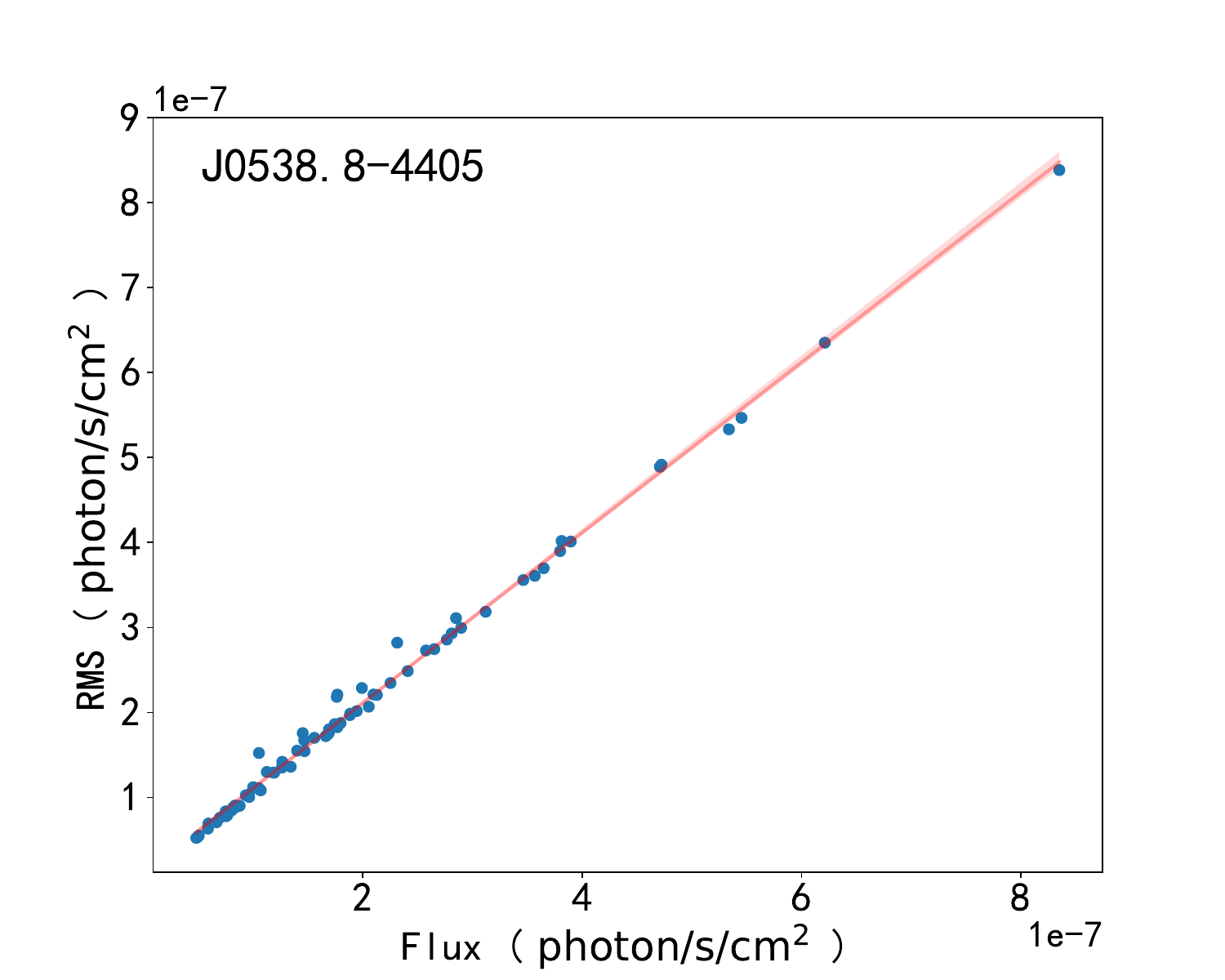}
  \end{minipage}
  \begin{minipage}{0.32\textwidth}
    \centering
    \includegraphics[width=\linewidth]{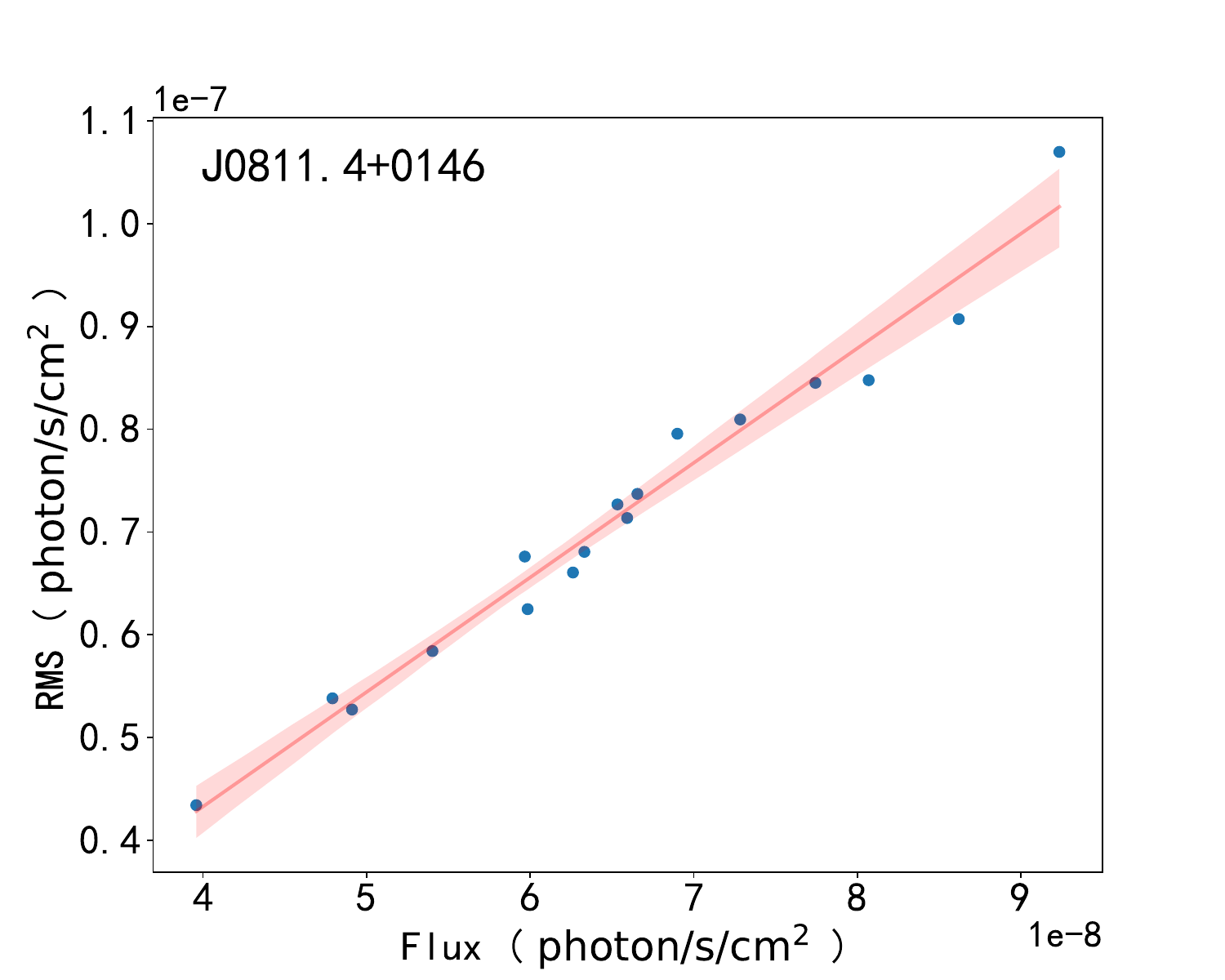}
  \end{minipage}
  \hfill
  \begin{minipage}{0.32\textwidth}
    \centering
    \includegraphics[width=\linewidth]{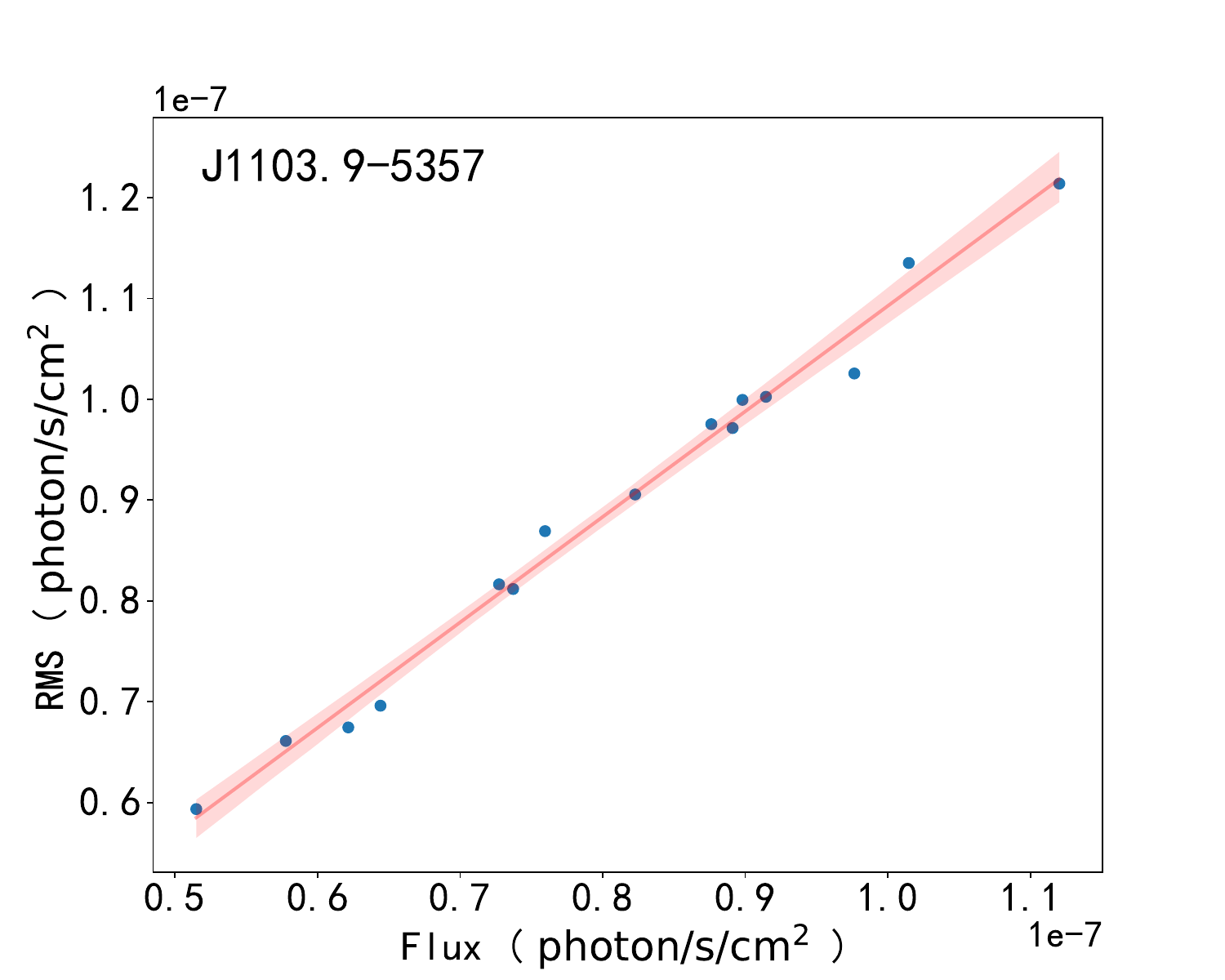}
  \end{minipage}
  \hfill
  \begin{minipage}{0.32\textwidth}
    \centering
    \includegraphics[width=\linewidth]{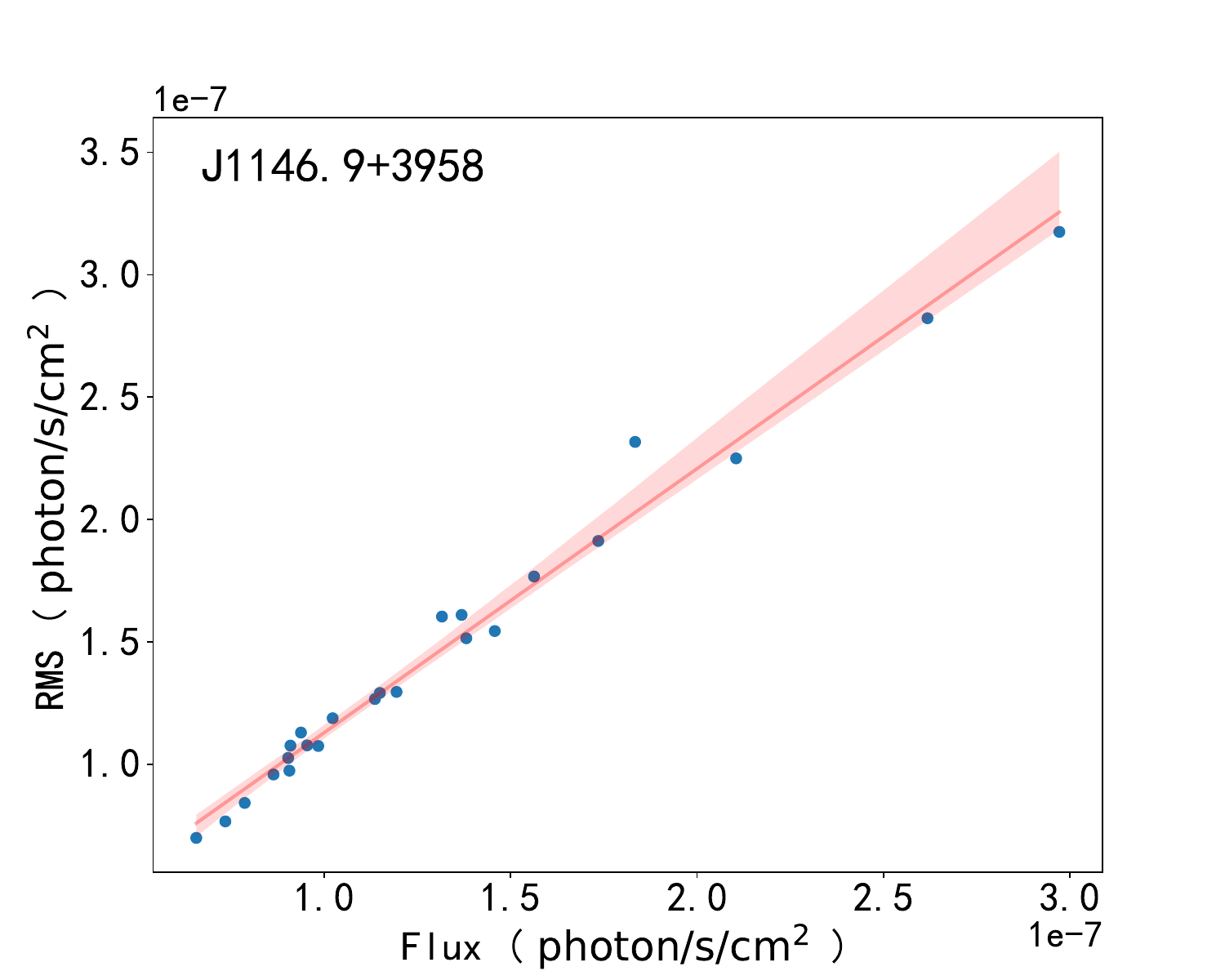}
  \end{minipage}
  \begin{minipage}{0.32\textwidth}
    \centering
    \includegraphics[width=\linewidth]{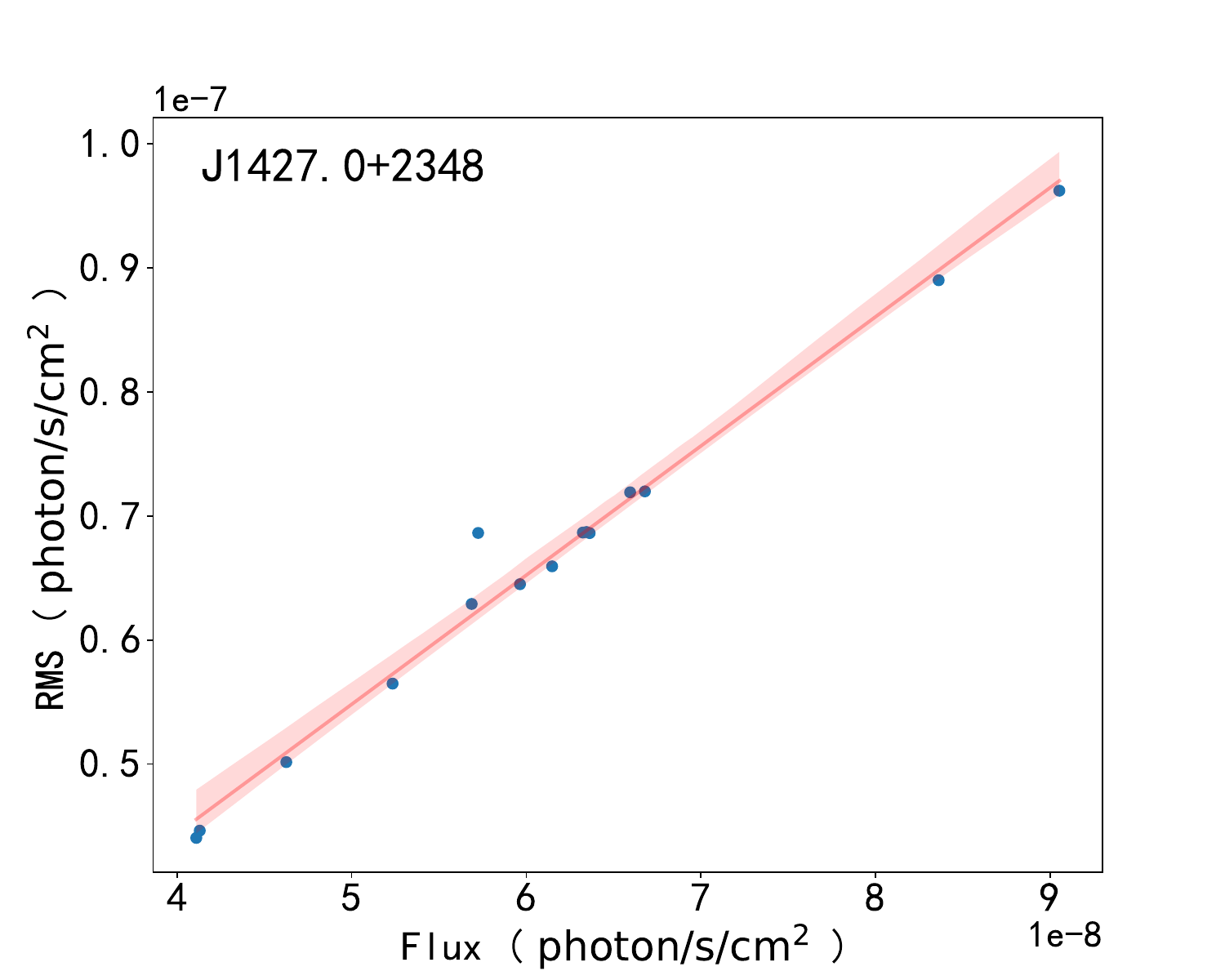}
  \end{minipage}
  \hfill
  \begin{minipage}{0.32\textwidth}
    \centering
    \includegraphics[width=\linewidth]{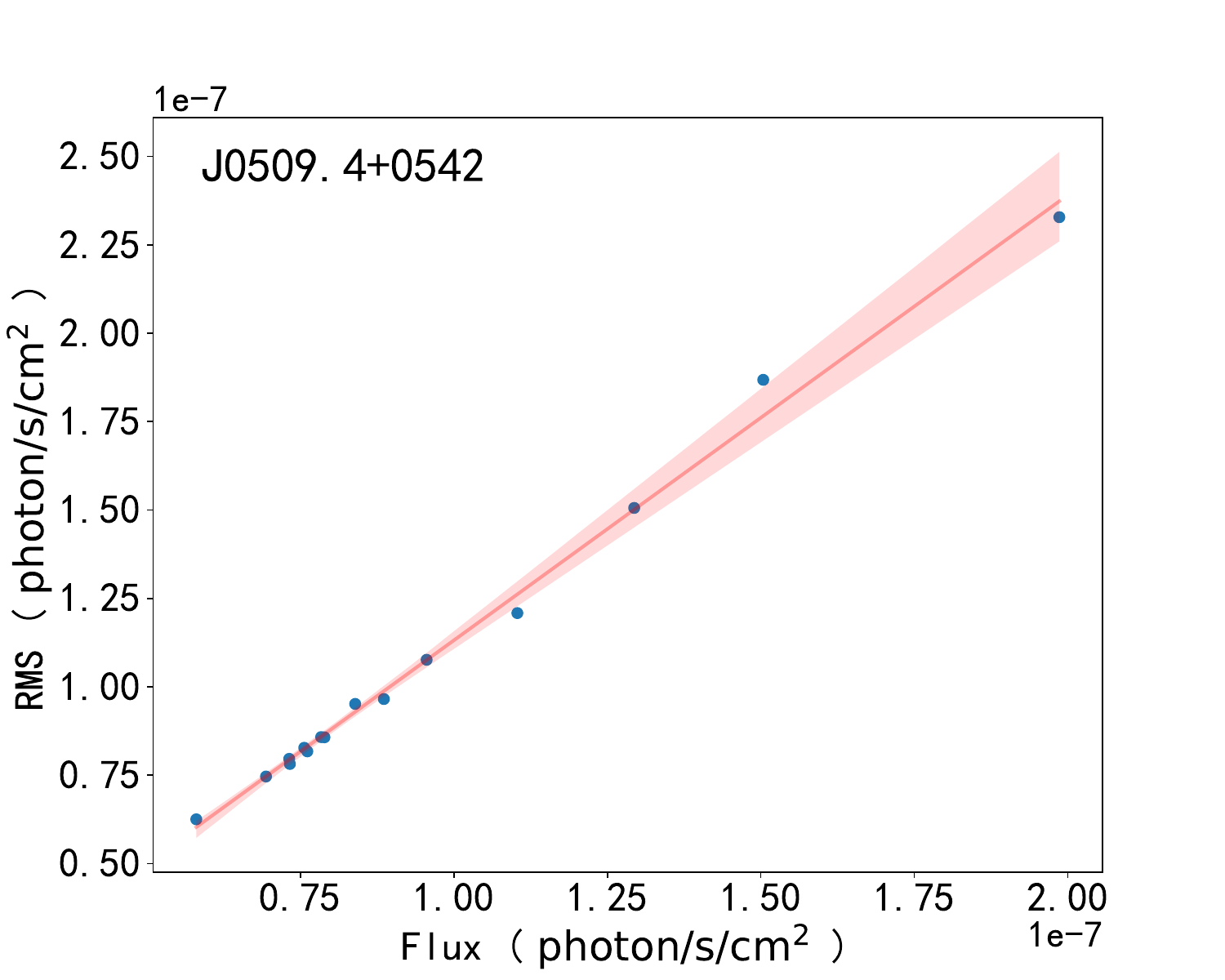}
  \end{minipage}
  \hfil
  \begin{minipage}{0.32\textwidth}
    \centering
    \includegraphics[width=\linewidth]{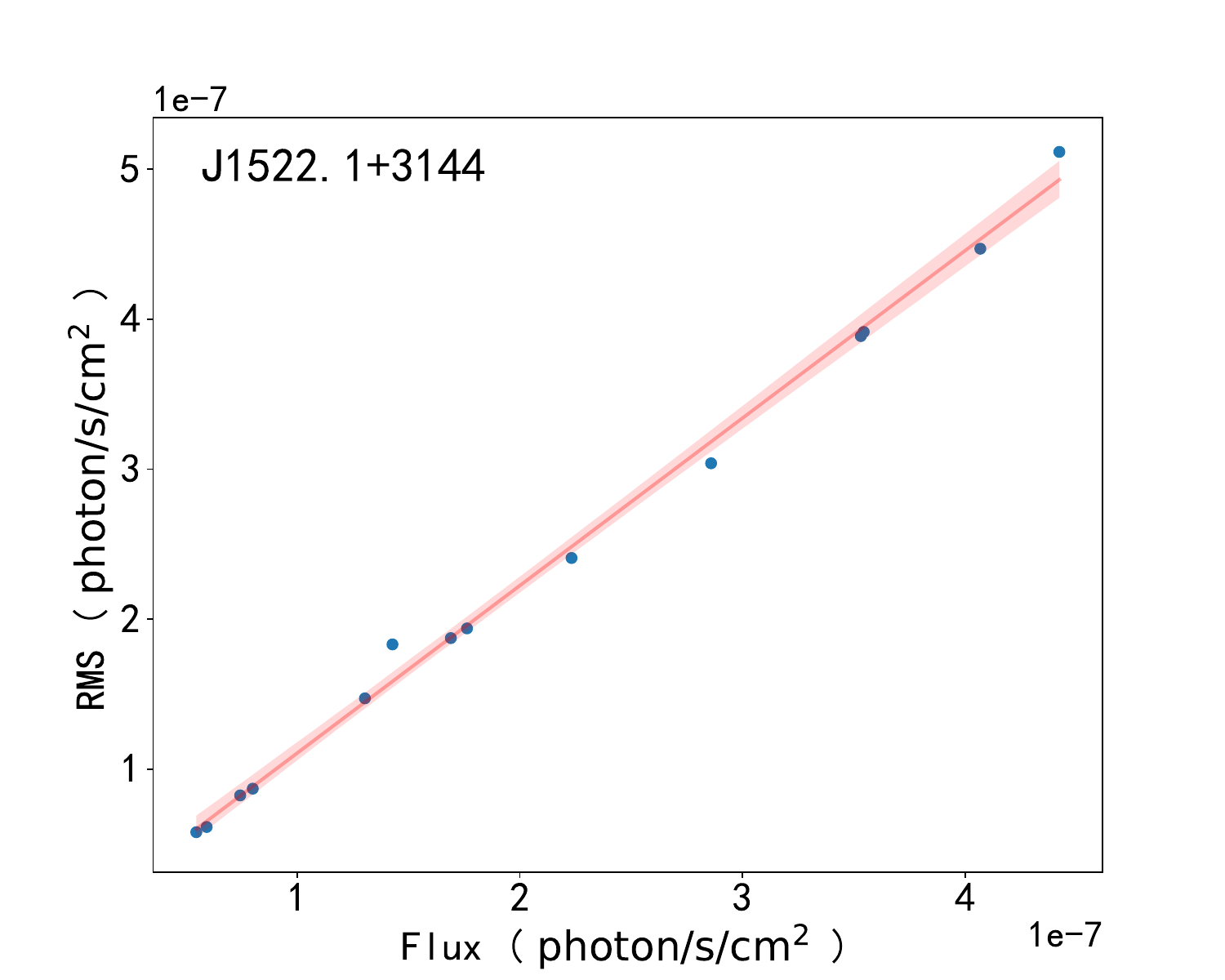}
  \end{minipage}
  \caption{Graphs of RMS-Flux relation of 3 randomly selected sources for each binnigs. The first row is the 20 data points per bin for RMS-Flux relation. The second row is the 40 data points per bin for RMS-Flux relation. The third row is the 365 days per bin for RMS-Flux relation. The figure lists only the first nine 9 sources. The entire figure is available in its entirety in machine-readable form.}
\label{Fig3}
\end{figure}
The RMS-Flux relation has been most convincingly demonstrated on relatively short timescales in X-ray band. In order to study RMS-Flux relations of blazars in $\gamma$-ray band, the 1414 blazars are choosed from Fermi-LAT LCR.
In order to avoid the too small number of bins that will affect the statistical significance of fitting results, we choose blazars with data points greater than 200 as a subsample.
After removing sources with less than 200 data points from this large sample, we obtain a sample including 41 BCUs, 236 BL Lacs, and 272 FSRQs.
We separately fit and estimate the RMS-Flux relation for three different binnings of light curve: 20 data points, 40 data points, and one year (365 days) per bin, and calculate their Pearson correlation coefficients. This three kinds of binnings are based on time series.
The first two binnings involve dividing the data point into bin of fixed point numbers, with each bin consisting of 20 data points and 40 data points, respectively. In order to further investigate whether the binnings has an impact on the RMS-Flux
relation, the third binnings involves grouping the data based on fixed time intervals, with all points within 365 days (yearly) interval forming a bin.
We calculate the RMS and average flux for each bin and perform a linear regression analysis. Therefore, we obtain the Pearson correlation coefficient for the RMS-Flux. We can see that the interdependence between RMS and flux is noteworthy, with a substantial and coherent correlation between the two.
Linear fitted plot of RMS-Flux relation are shown in Figure~\ref{Fig3}.

\begin{table}[H]
\fontsize{6.5}{8}\selectfont
\caption{Statistical results for three different binnings, including 20 data points,40 data points and 365 days. The table lists only the first 9 blazars. This entire table is available in its machine-readable form.}\label{tab3}
\setlength{\tabcolsep}{0.035cm}
\begin{tabular}{cccccccccccccccc}
\hline\hline
\multirow{2}*{Name(type)} &\multicolumn{3}{c}{{20 data points bin}}& {} &\multicolumn{3}{c}{{40 data points bin}}& {} & &\multicolumn{3}{c}{{365 days}}\\
\cline{2-4}\cline{6-8}\cline{10-12}
 & Slope& Intercept & Pearson correlation & & Slope& Intercept & Pearson correlation & & Slope& Intercept & Pearson correlation & & \\
\hline
J0001.5+2113(fsrq) & 1.267 & $-1.983\times10^{-8}$ & 0.987 & & 1.246 & $-1.027\times10^{-8}$& 0.988 & & 
1.433 & $-4.117\times10^{-8}$ & 0.994 \\
J0009.3+5030(bll) & 4.363 & $-1.276\times10^{-7}$ & 1.000 & & 6.247 & $-2.037\times10^{-7}$ & 1.000 & & 4.675&$7.123\times10^{-8}$ & 0.990 \\
J0011.4+0057(fsrq) & 1.357 &$-1.514\times10^{-8}$ & 0.961 & & 1.298 &$-6.288\times10^{-9}$ & 0.991 & &
1.143 & $9.950\times10^{-9}$  & 0.941 \\
J0017.5-0514(fsrq) & 1.462 & $-2.569\times10^{-8}$ & 0.786 & & 1.649 &$-5.104\times10^{-8}$  & 0.957 & &
1.434 & $-3.059\times10^{-8}$ & 0.964 \\
J0019.6+7327(fsrq) & 1.390 & $-4.960\times10^{-8}$ & 0.962 & & 1.304 & $-3.301\times10^{-8}$ & 0.977 & &
1.345 & $-4.573\times10^{-8}$ & 0.985 \\
J0022.5+0608(bll) & 1.096 & $-2.997\times10^{-10}$ & 0.994 & & 1.076 & $1.207\times10^{-9}$  & 0.995 & & 
1.005 & $5.024\times10^{-9}$  & 0.980 \\
J0028.4+2001(fsrq) & 1.040 & $1.328\times10^{-8}$  & 0.997 & & 1.038 & $2.068\times10^{-8}$ & 0.998 & &
1.263 & $-1.431\times10^{-8}$ & 0.998 \\
J0030.3-4224(fsrq) & 1.432 & $-2.576\times10^{-8}$ & 0.980 & & 1.651 & $-4.226\times10^{-8}$ & 0.978 & &
1.164 & $2.797\times10^{-9}$ & 0.894 \\
J0030.6-0212(bcu) & 1.159 & $-1.776\times10^{-9}$ & 0.983 & & 1.146 & $-4.479\times10^{-10}$ & 0.991 & &
1.326 & $-1.904\times10^{-8}$  & 0.995\\
\hline
\end{tabular}\\
\end{table}

\begin{table}[H]
\centering
\begin{minipage}[]{\textwidth}
\caption[]{Statistical analysis of the slope and intercept for the RMS-Flux relation.
\label{tab4}}\end{minipage}
\setlength{\tabcolsep}{1pt}
\small
\begin{tabularx}{0.95\textwidth}
{c@{\hspace{1em}}c@{\hspace{1em}}c@{\hspace{1em}}c@{\hspace{1em}}c@{\hspace{1em}}c}
    \hline\noalign{\smallskip}
\multicolumn{6}{c}{ rms = m*Flux+c}                                       \\
  \hline\noalign{\smallskip}
points   & mean of m & mean of c & the fraction of m\textgreater{}0 & the fraction of   c\textgreater{}0 & Pearson correlation \\
  \hline\noalign{\smallskip}
\multicolumn{6}{c}{20 data points}                                       \\
  \hline\noalign{\smallskip}
FSRQs      & 1.860  & $-1.190\times10^{-8}$ & 100.00\%                                 & 26.84\%                                       & 0.977              \\
BL Lacs & 1.996 & $-4.159\times10^{-10}$ & 100.00\%                                & 29.24\%                                       & 0.961              \\
BCUs       & 1.985 & $-5.483\times10^{-8}$ & 100.00\%                                & 26.83\%                                       & 0.979              \\
  \hline\noalign{\smallskip}
\multicolumn{6}{c}{40 data points}                                       \\
  \hline\noalign{\smallskip}
FSRQs      & 2.197 & $-6.282\times10^{-8}$ & 100.00\%                                 & 22.06\%                                       & 0.977              \\
BL Lacs & 2.417 & $-3.587\times10^{-8}$ & 100.00\%                                & 28.39\%                                       & 0.953              \\
BCUs       & 2.453 & $-1.198\times10^{-7}$ & 100.00\%                                & 24.39\%                                         & 0.977              \\
  \hline\noalign{\smallskip}
\multicolumn{6}{c}{365 days}                                       \\
  \hline\noalign{\smallskip}
FSRQs      & 2.032 & $1.232\times10^{-7}$  & 100.00\%                                & 17.28\%                                          & 0.981              \\
BL Lacs & 2.328 & $1.128\times10^{-8}$  & 100.00\%                                & 19.49\%                                          & 0.970              \\
BCUs       & 1.966 & $7.056\times10^{-8}$  & 100.00\%                                & 17.07\%                                            & 0.980              \\
  \noalign{\smallskip}\hline
\end{tabularx}
\end{table}

\begin{figure}[H]
  \centering
  \begin{minipage}{0.32\textwidth}
    \centering
    \includegraphics[width=4.8cm]{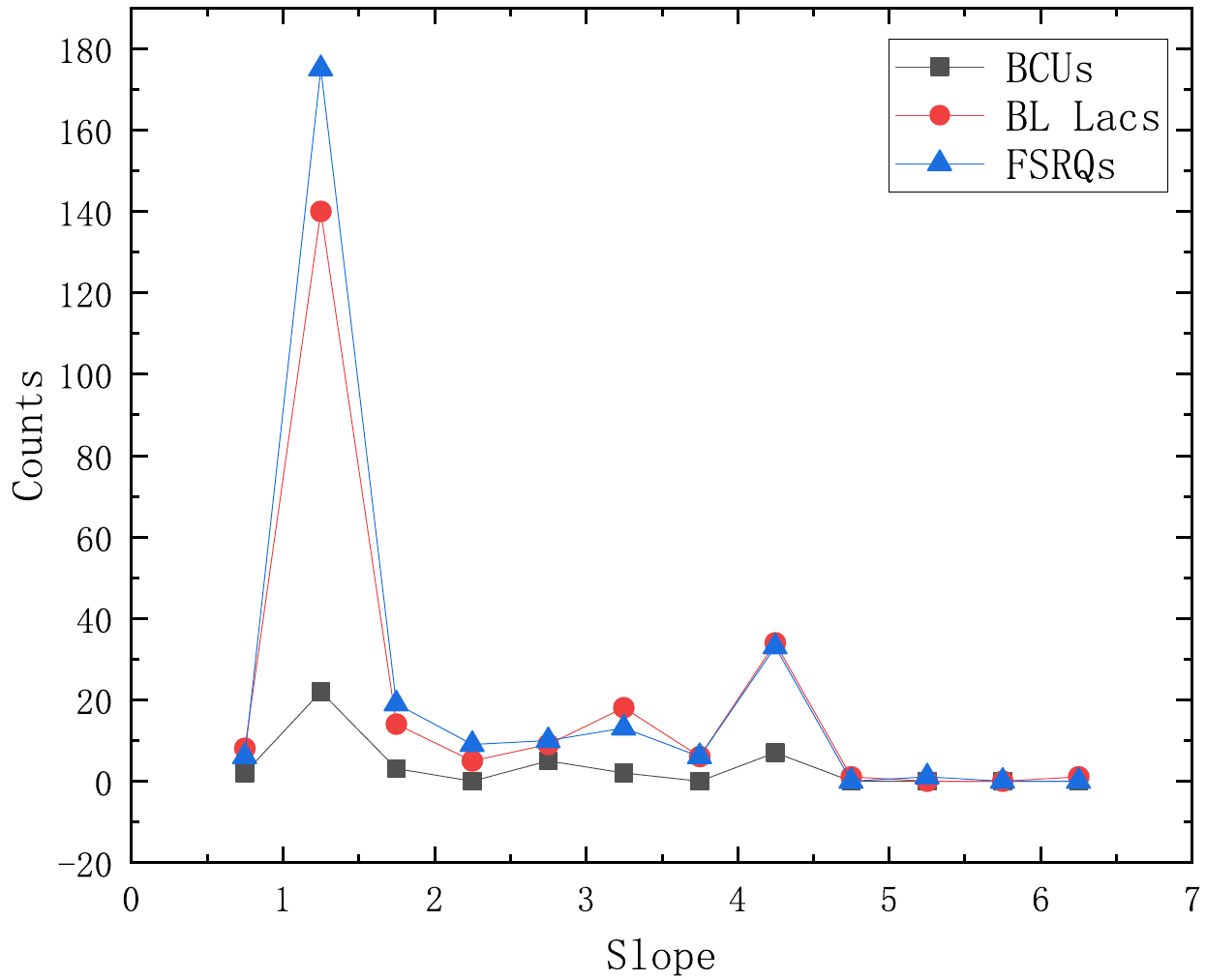}
  \end{minipage}
  \hfill
  \begin{minipage}{0.32\textwidth}
    \centering
    \includegraphics[width=4.8cm]{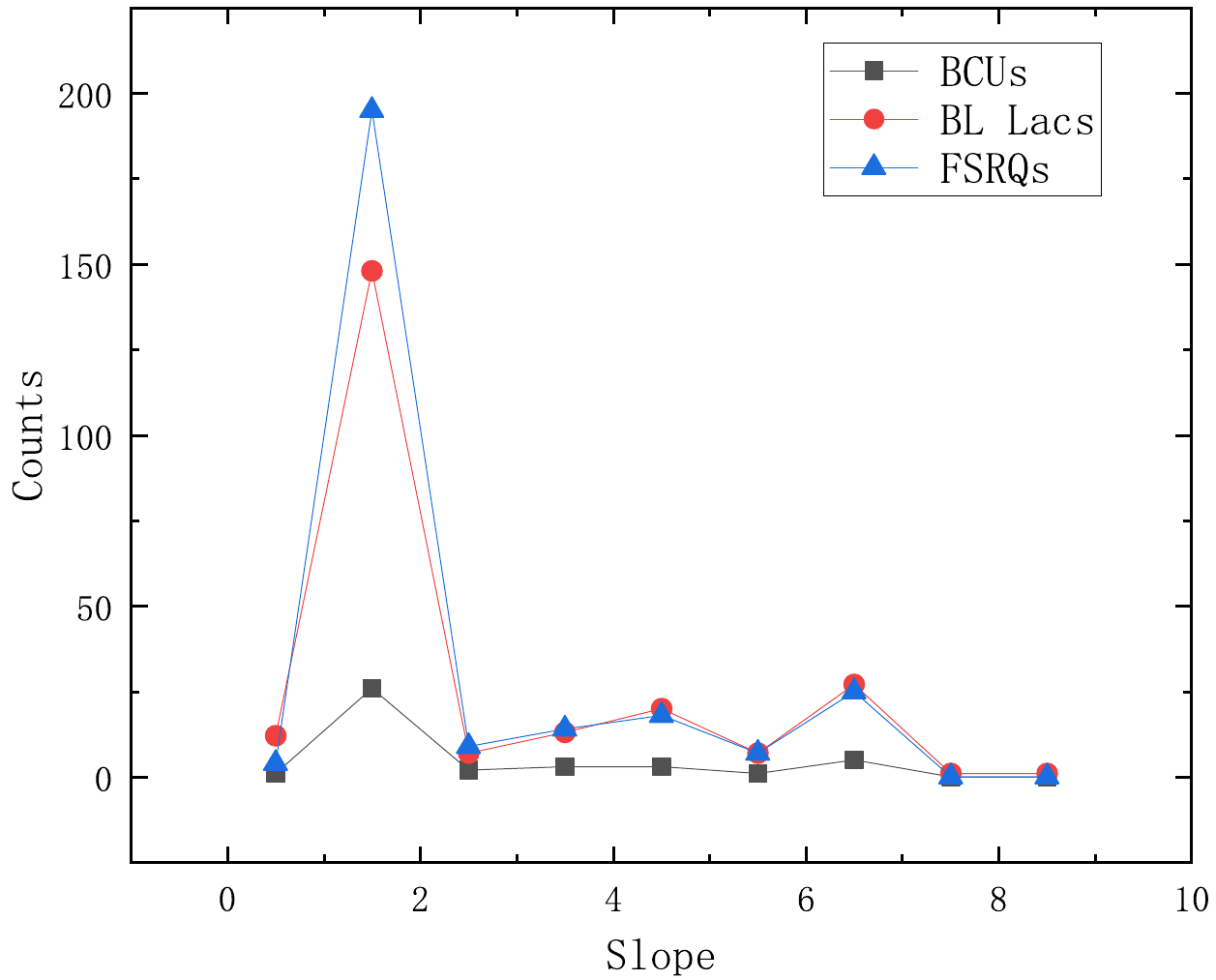}
  \end{minipage}
  \hfill
  \begin{minipage}{0.32\textwidth}
    \centering
    \includegraphics[width=4.8cm]{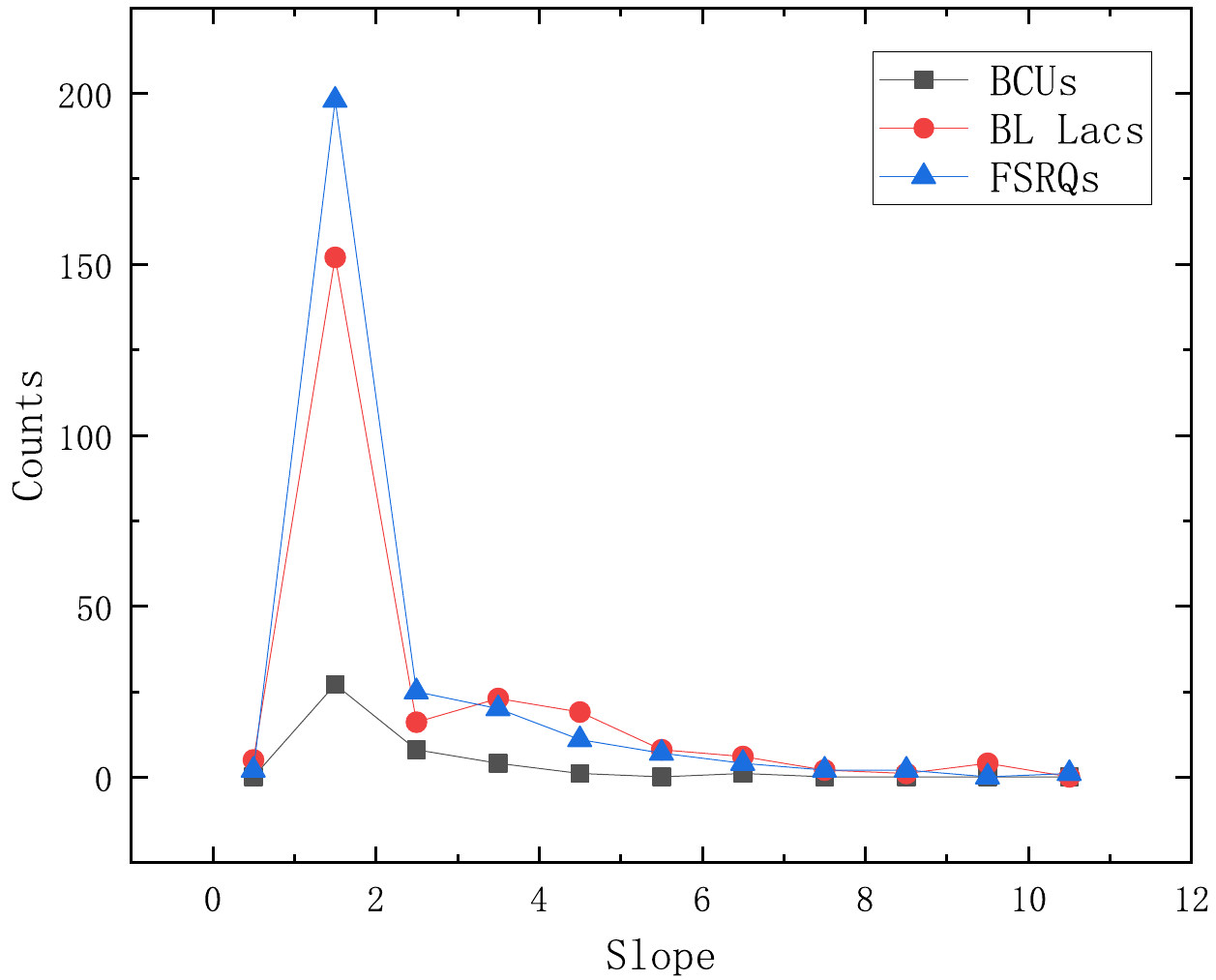}
  \end{minipage}
  \begin{minipage}{0.32\textwidth}
    \centering
    \includegraphics[width=\linewidth]{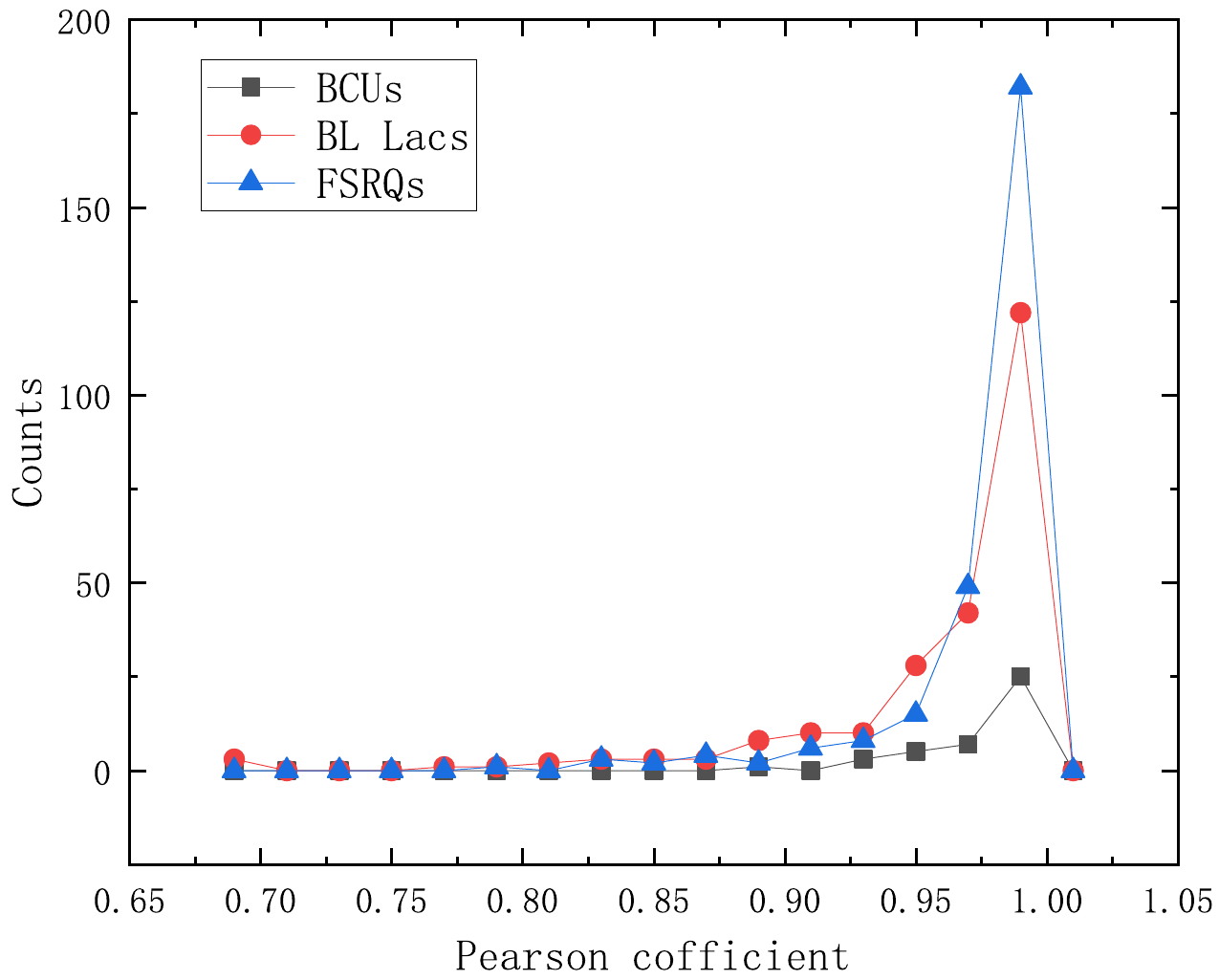}
    \caption*{(a) 20 data points}
  \end{minipage}
  \hfill
  \begin{minipage}{0.32\textwidth}
    \centering
    \includegraphics[width=\linewidth]{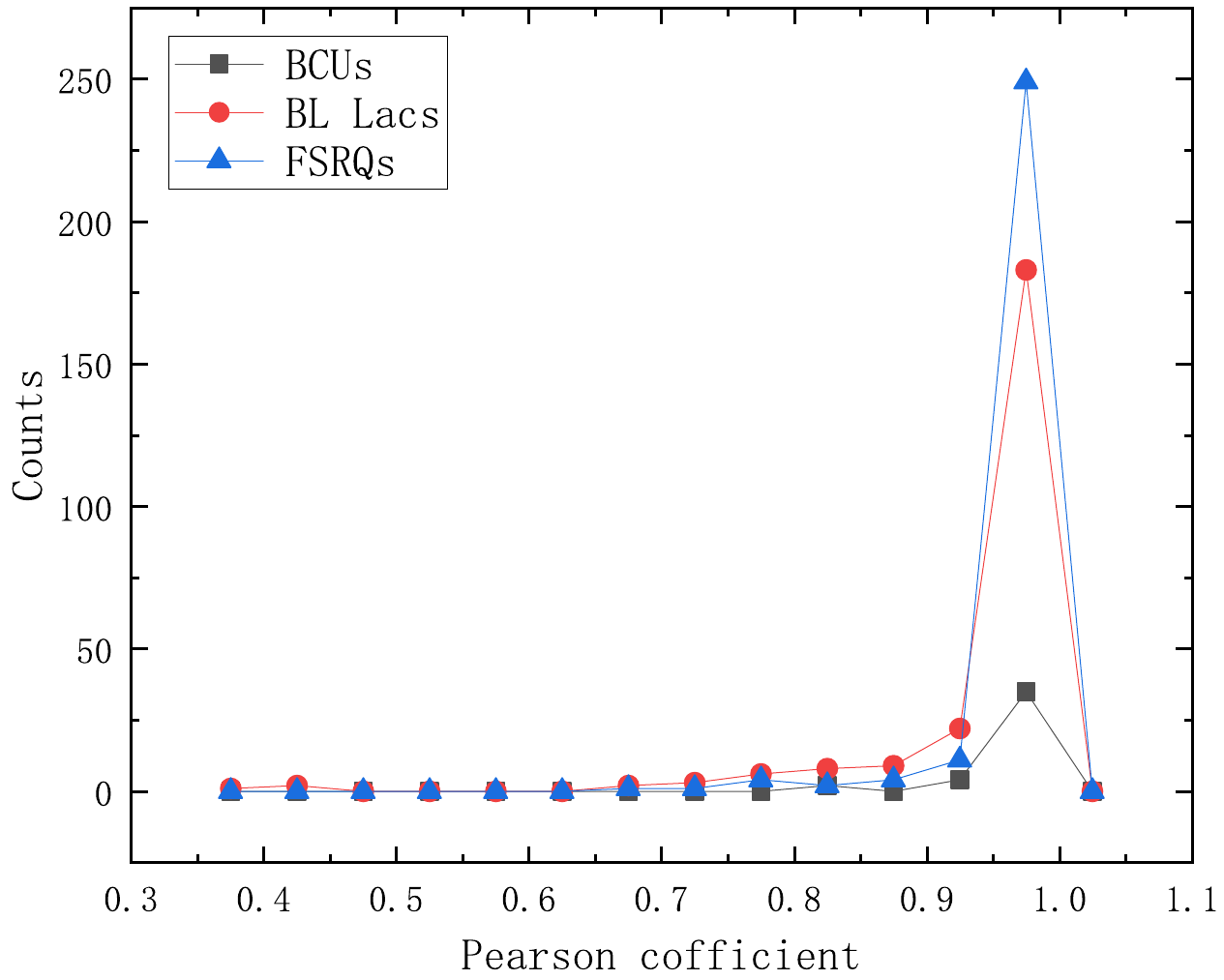}
    \caption*{(b) 40 data points}
  \end{minipage}
  \hfill
  \begin{minipage}{0.32\textwidth}
    \centering
    \includegraphics[width=\linewidth]{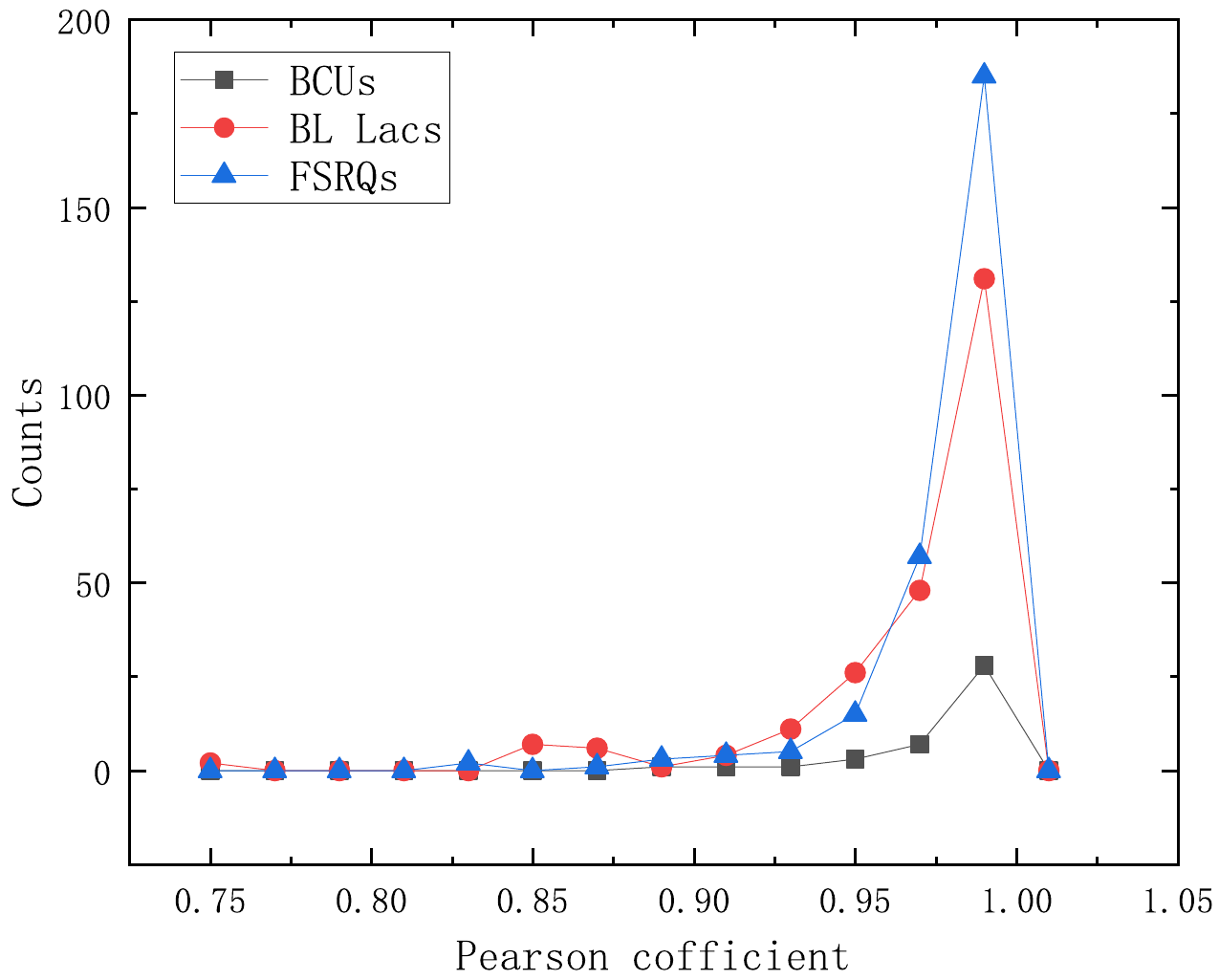}
    \caption*{(c) 365 days}
  \end{minipage}
  \caption{Statistical results for three different binnings, including 20 data points, 40 data points and 365 days, according to the column a, b, c, respectively. The first row is the distribution of slope for three binnings. The second row is the distribution of Pearson correlation coefficients.}
\label{Fig4}
\end{figure}

Statistical results of three different binnings are shown in Figure~\ref{Fig4} and Table~\ref{tab3}.
As shown in Table~\ref{tab4}, when 20 data points are taken as a bin, the mean slope of the RMS-Flux for FSRQs, BL Lacs, and BCUs are 1.860, 1.996, and 1.985, respectively.
The mean Pearson correlation coefficients for RMS-Flux relation for all three types are 0.977, 0.961, and 0.979, respectively.
Thus, there is a strong correlation between RMS and flux.
For the other two binnings, the Pearson correlation coefficients for RMS-Flux are $\sim1$, indicating strong liner RMS-Flux relation for all 549 sources, as shown in the Table~\ref{tab4}.

Based on the fitting of the RMS-Flux relation for BL Lacs, FSRQs, and BCUs using different grouping schemes, the slopes for these three types of sources are greater than zero. The BL Lacs exhibit a steeper slope in relation to the FSRQs. However, the differences of slopes obtained from three types of sources are too small, therefore the slope cannot be used as a criterion to distinguish between the three source types.

\section{DISCUSSION AND CONCLUSION}
\label{sect:discussion}
\begin{figure}[H]
  \centering
  \begin{minipage}{0.32\textwidth}
    \centering
    \includegraphics[width=5cm]{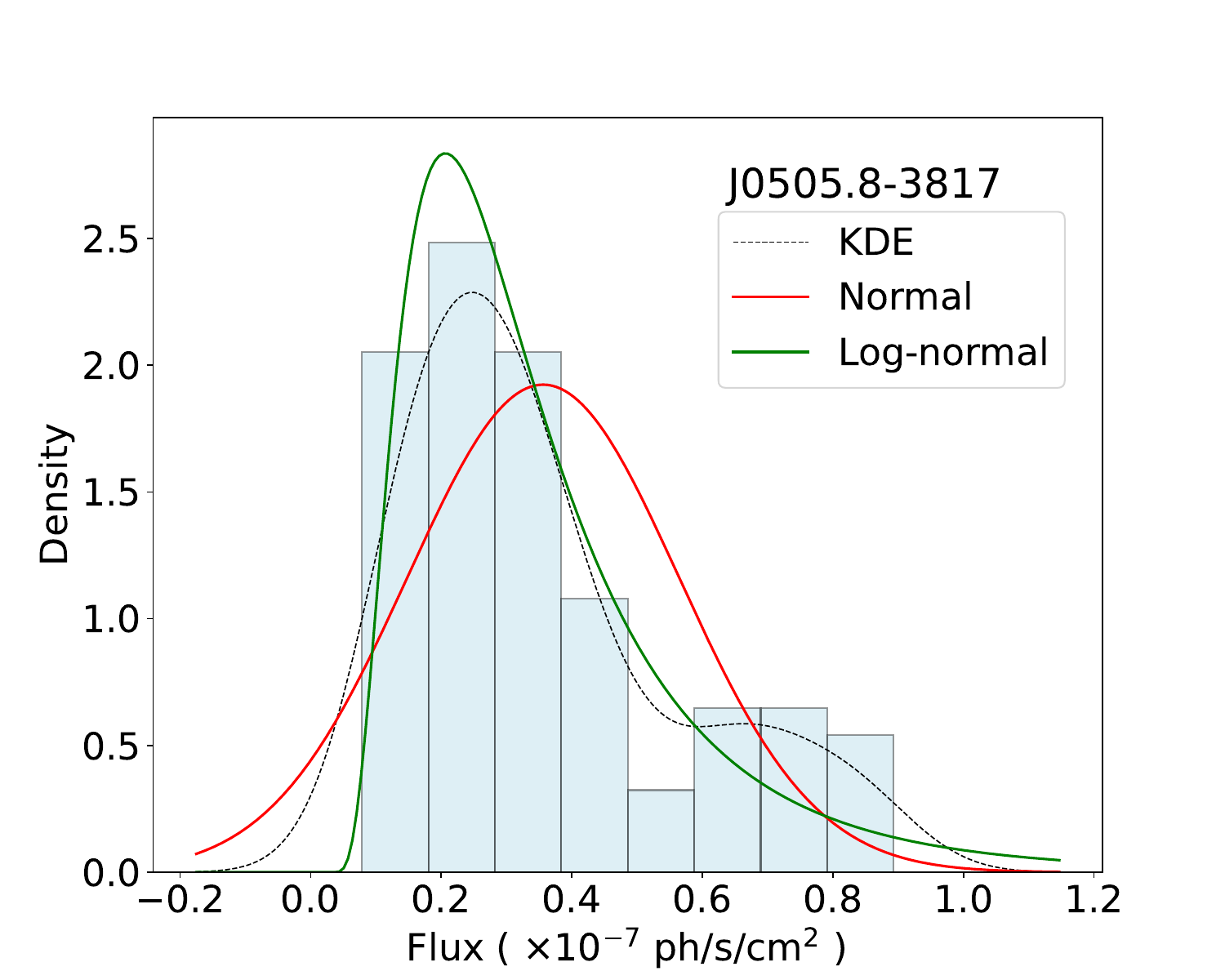}
  \end{minipage}
  \hfill
  \begin{minipage}{0.32\textwidth}
    \centering
    \includegraphics[width=5cm]{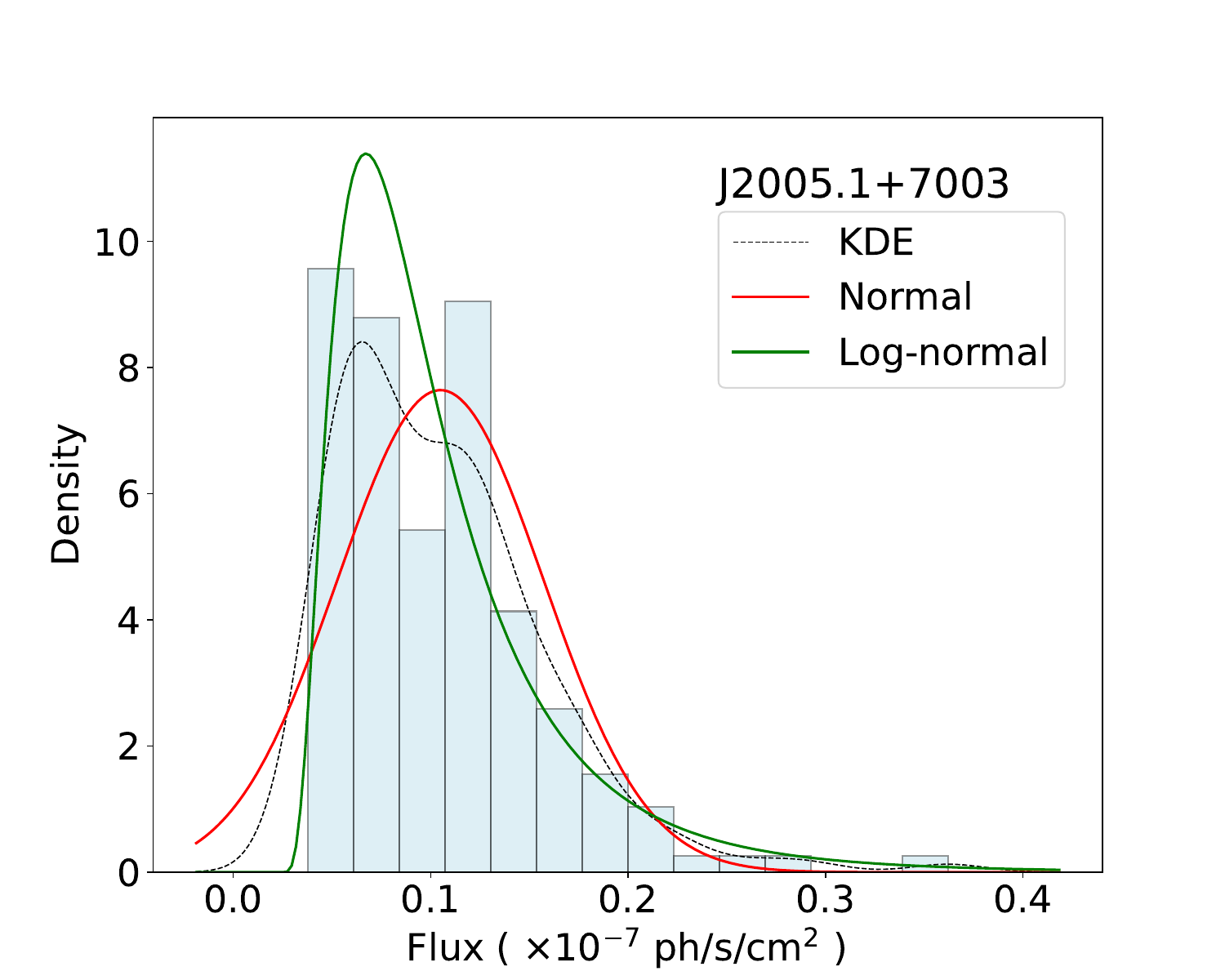}
  \end{minipage}
  \hfill
  \begin{minipage}{0.32\textwidth}
    \centering
    \includegraphics[width=5cm]{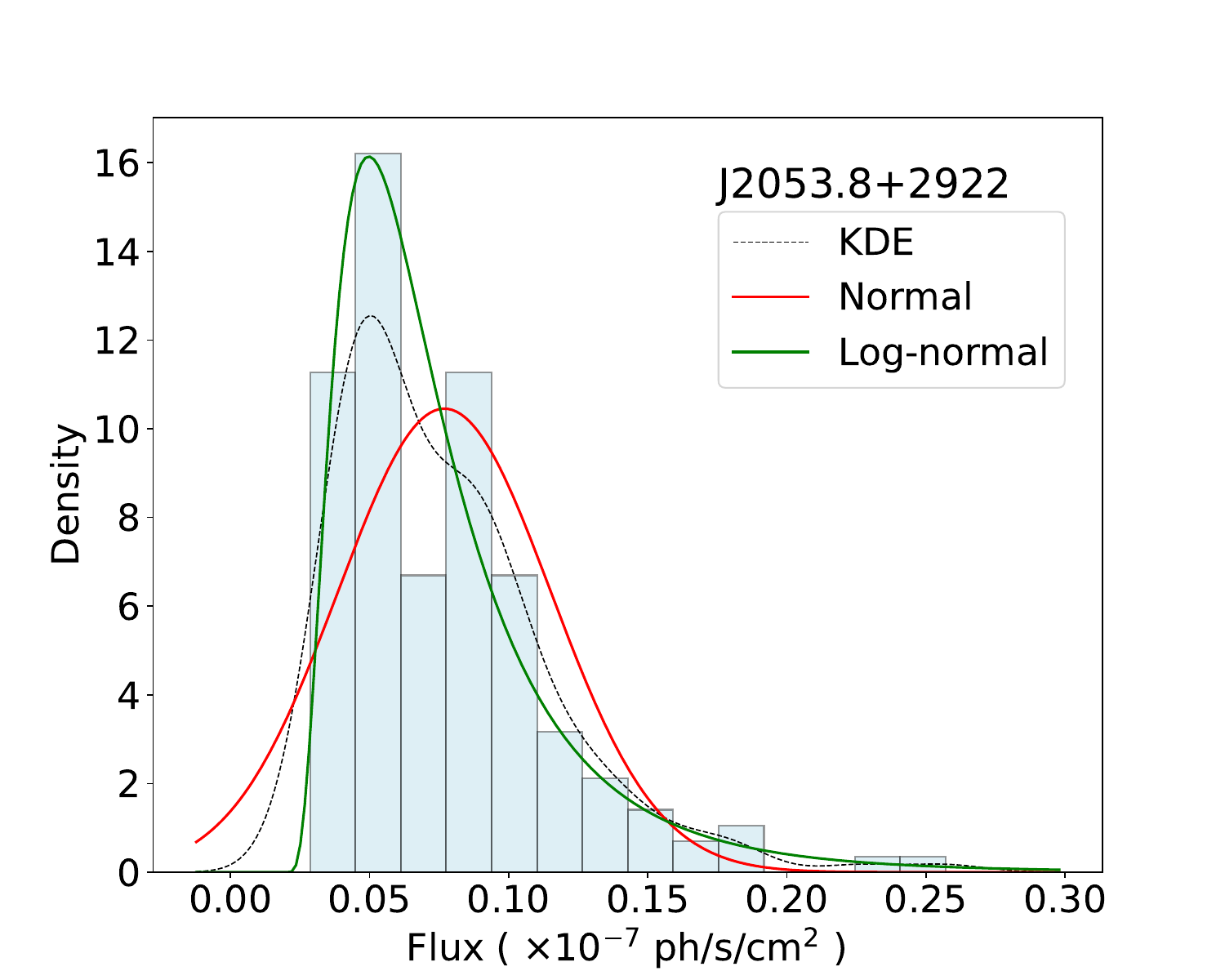}
  \end{minipage}
  \begin{minipage}{0.32\textwidth}
    \centering
    \includegraphics[width=5cm]{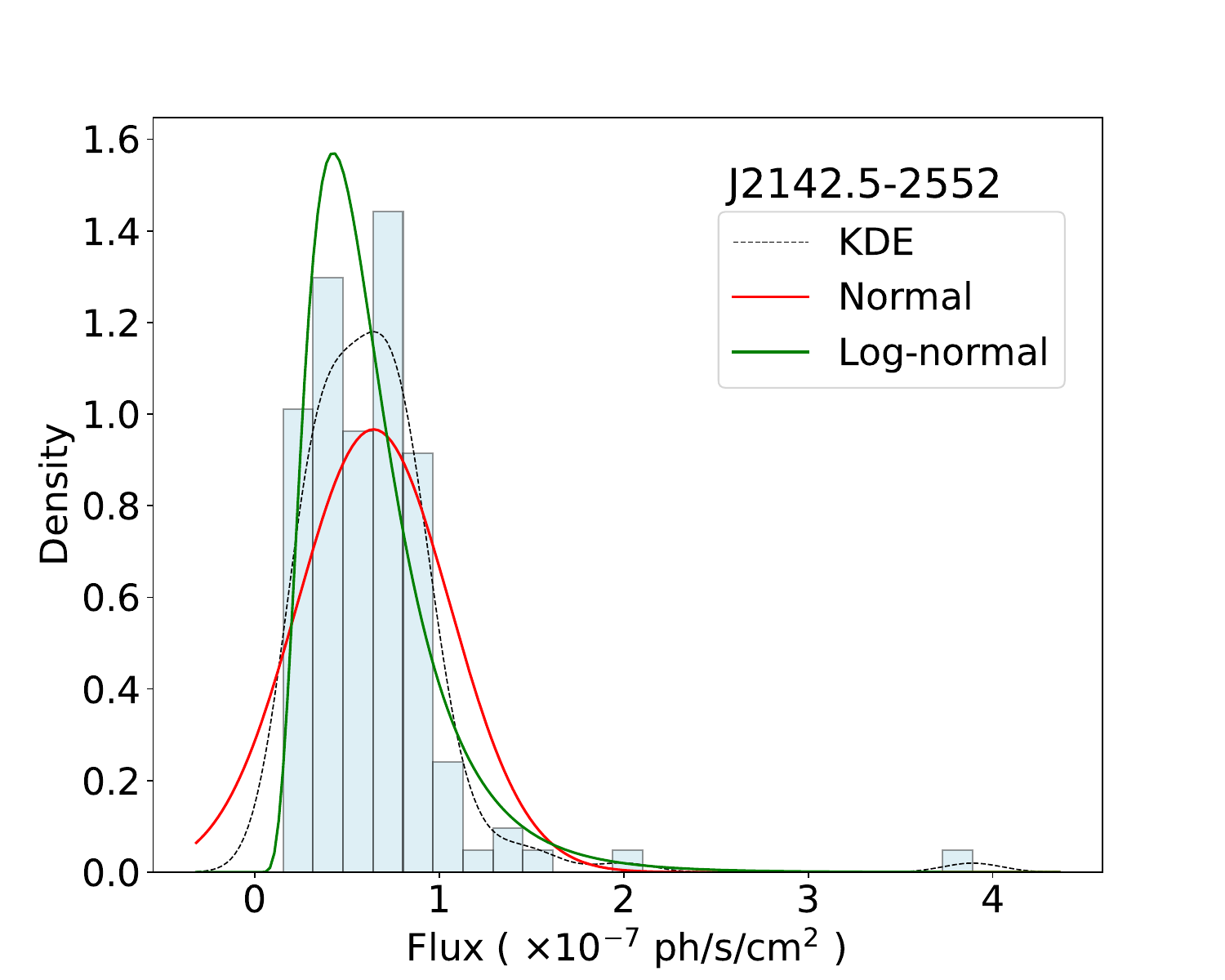}
  \end{minipage}
  \hfill
  \begin{minipage}{0.32\textwidth}
    \centering
    \includegraphics[width=5cm]{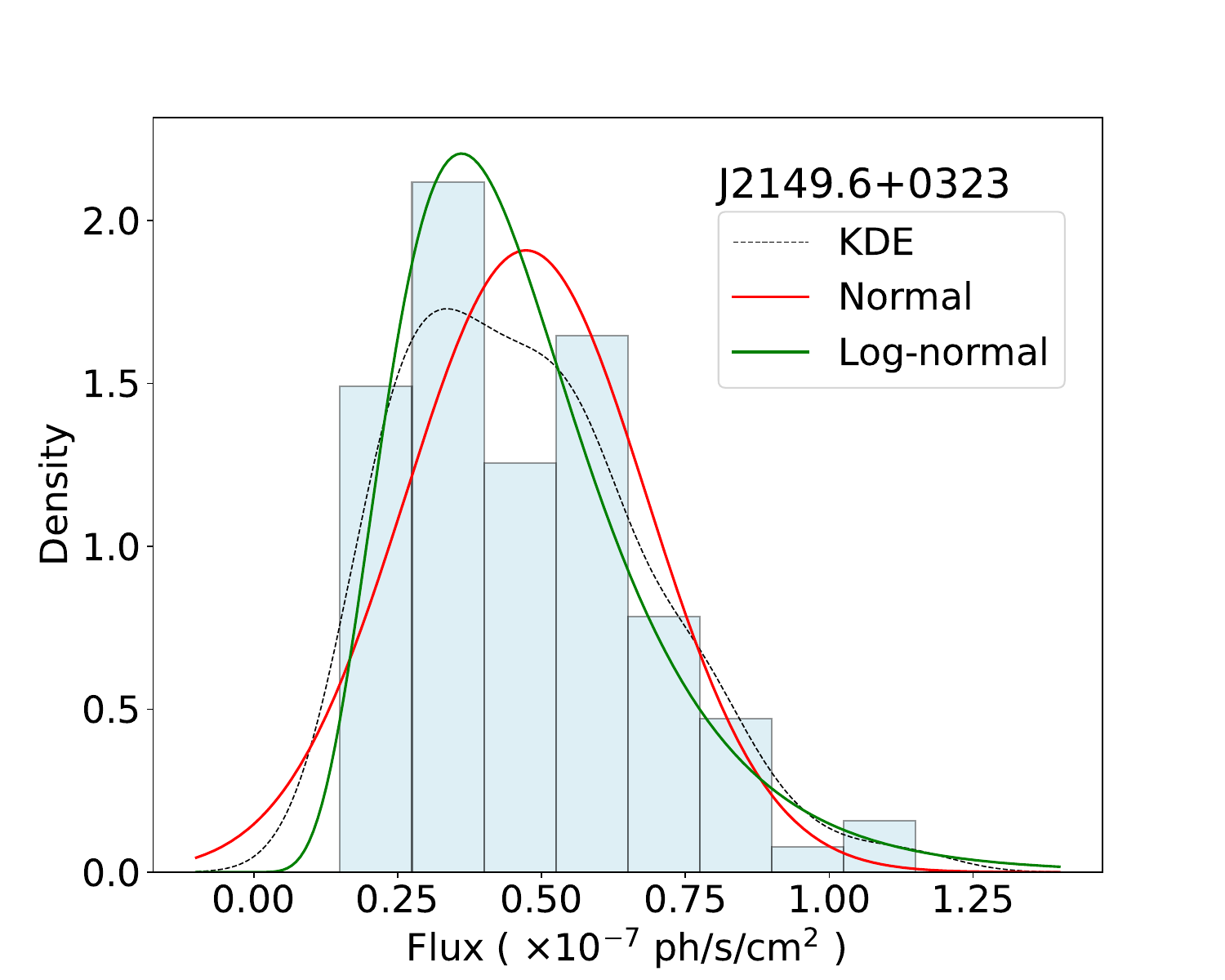}
  \end{minipage}
  \hfill
  \begin{minipage}{0.32\textwidth}
    \centering
    \includegraphics[width=5cm]{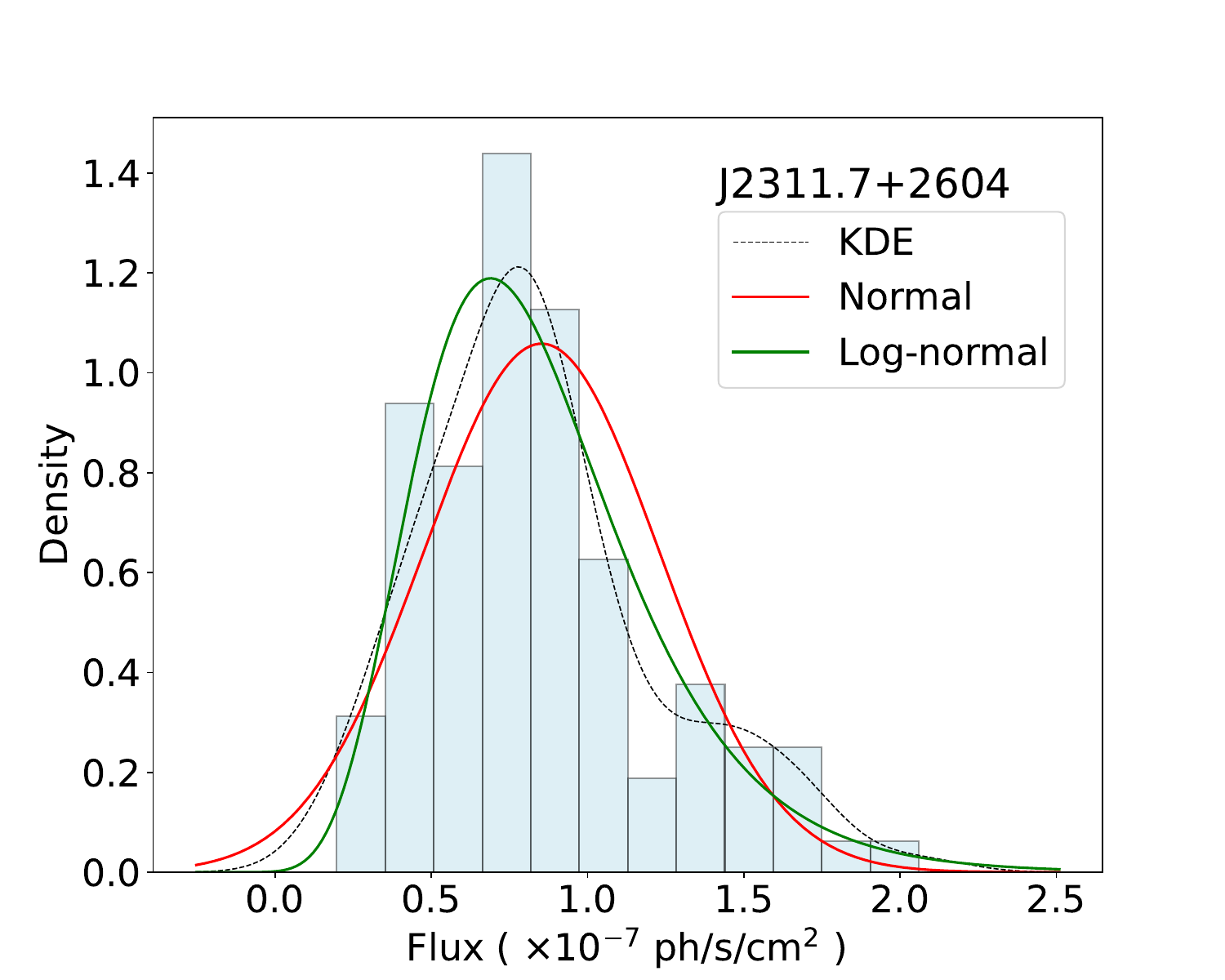}
  \end{minipage}
  \caption{Bimodal of the $\gamma$-ray flux distribution (hatched blue), and normal (red) and log-normal (green) PDF fitting to the histogram  for the 6 blazars, the black dotted line represents the kernel density estimation (KDE) curve. Since the statistics is not significant enough, these fits should be taken only as indicative.}
\label{Fig5}
\end{figure}

In this work, we investigate the variability properties of the $\gamma$-ray by analyzing the flux distribution and the linear RMS-Flux relation of a large blazars samples from Fermi-LAT LCR, including 572 FSRQs, 477 BL Lacs, and 365 BCUs.
We fitted the flux distribution histograms to determine whether the flux distribution conforming to normal or log-normal functions.
Of the 572 FSRQs, 199 FSRQs consist with a log-normal distribution for flux and only 9 FSRQs follow a normal distribution.
Of the 477 BL Lacs, 230 BL Lacs consist with a log-normal distribution and only 4 BL Lacs follow a normal distribution.
In the case of 365 BCUs, there are 179 BCUs consist with log-normal distribution of flux and only 5 BCUs consist with normal. The probability that the flux distribution of BL Lacs shows log-normal is greater than that of FSRQs.
Interestingly, we find that the flux of a few of sources is normally distributed, and the flux distribution is also consist with log-normal distribution. While further analysis is required for those blazars whose distribution does not conform to either the log-normal distribution or the normal distribution.
In particular, the S5 0716+714 $\gamma$-ray flux distribution with both normal and log-normal PDF were fitted and found that the normal PDF fitted light curves (28-days binnings) best (\citealt{Penil+etal+2022}). However, for the same source, our fit to the flux distribution follow a log-normal distribution (3-days binnings). \cite{Penil+etal+2022} used the 28-days binning for the light curves, while we use 3-days binned light curves. This indicate binning for light curves will affect the fitting results.
As shown in Figure~\ref{Fig5}, we also found some histograms showing bimodal patterns, and the statistics of this distribution is not
significant enough, hence these result should be taken only as an indicative.
In addition to using Fermi data with TS\textgreater{4}, we try to use the Fermi data with TS\textgreater{9} to study flux distribution and summarize the results that under the K-S, S-W, and Normality tests. 
We conduct statistical analysis on the data of the three test results. For the K-S test, 86.21\% of the blazars of flux distributions exist a log-normal distribution, while 57.99\% of the blazars of flux distributions show a normal distribution. In the S-W test, 61.60\% of the blazars of flux distributions exist a log-normal distribution, while 24.47\% of the blazars of flux distributions show a normal distribution. For the Normality test, 60.82\% of the blazars of flux distributions exist a log-normal distribution, while 28.08\% of the blazars of flux distributions show a normal distribution.
The flux distribution also generally tends to be log-normal. The number of data point with TS\textgreater{9} per bins is too small that will affect the statistical significance of fitting results. So we have not analyze the RMS-Flux relationship of the the data with TS\textgreater{9}.

There is a lot of literatures that explain why the log-normal distribution in many source and in different bands (\citealt{ Kirk+etal+1998, Kushwaha+etal+2017, Shah+etal+2018, Shah+etal+2020, Giebels+Degrange+2009, Khatoon+etal+2020, Scargle+2020}).
The variability of the blazars may be due to a combination of intrinsic events of the source, such as instability of both the disk and the jet, and extrinsic influences on the source's geometry and projection.
For instance, there are mainly three explanatory models for the log-normal distribution:
(1) In X-ray band, it was initially believed that the log-normal distribution of the flux was from accretion disks, the propagating fluctuations model proposed by Lyubarskii to explain this behavior (\citealt{Lyubarskii+1997, Meyer+2019, Khatoon+etal+2020};
(2) Inherent particle acceleration in the acceleration region on acceleration timescales leads to linear normal perturbations, which result in log-normal distributions of the flux (\citealt{Narayan+etal+2012, Giannios+etal+2009, Sinha+etal+2018});
(3) In addition, the log-normal distribution of blazars has also been interpreted as the sum of emission of small jets, i.e., the “minijets in a jet” statistical model (\citealt{Biteau+Giebels+2012a, Biteau+Giebels+2012b}).
The “minijets in a jet” model was developed based on the second explanatory model. 
In the “minijets in a jet” model, emisson from identical, independent but randomly oriented minijets follows the Pareto distribution, which produce both normal and log-normal distributions, while preserving a linear RMS-Flux relationship in all cases (\citealt{Biteau+Giebels+2012b}).
The “minijets in a jet” model is favored by our statistical results: in our blazar sample, in addition to a lot of sources exhibiting log-normal distribution, some other sources also exhibit normal distribution.

In general, the log-normal distribution of the flux implies the linear RMS-Flux relation (\citealt{Biteau+Giebels+2012b}).
Furtherly, the liner RMS-Flux relation also implies that the variability process and the associated perturbations are multiplicative in nature (\citealt{Lyubarskii+1997, Arévalo+Uttley+2006}).
Mathematically, the addition process produces a normal distribution, while the multiplication process produces a log-normal distribution.
In X-ray band, the log-normal flux distribution and the linear RMS-Flux relation can be explained by the accretion disk, inward propagating fluctuations in the disk from different radii accumulate in a multiplicative way (\citealt{Arévalo+Uttley+2006}).
Whereas in $\gamma$-ray band, the variability of blazars is primarily associated with the non-thermal radiation from the jet. If the radiation is relativistically beamed, all possible contributing factors, such as such as the variable magnetic field and high-energy particle density, seed-photon density acted upon by the particle acceleration, and diffusion processes could be coupled in a complex manner resulting in the log-normal distribution of flux distribution in blazars(\citealt{Bhatta+Dhital+2020}).

Both normal and log-normal distributions can be interpreted as being special cases of a more general class of skewed distribution, such as Pareto distributions, with variable degrees of skewness. In Pareto distributions scenario, the resulting flux distribution has been
hold the RMS-Flux relation (\citealt{Biteau+Giebels+2012a}).
We also select the blazars with more than 200 data points for RMS-Flux fitting of the different types and binnings.
The fitting results show that the RMS and flux of three subtypes of blazars have a strong linear relationship in three kinds of bins, and their average Pearson coefficient is above 0.95. About 75\% of the blazars have an intercept less than 0. Besides, the results show that
the slope of BL Lacs was generally greater than that of FSRQs.
The strong correlation of the liner RMS-Flux relation of the blazars belongs to its intrinsic properties.
Among the 549 blazars with linear RMS-Flux relation and the flux distribution of 415, 183 and 187 blazars have lognormality based on the K-S, S-W, and Normality tests, respectively.
There are 172 sources are both collected in the 595 source sample and in the 549 source sample. Sources whose flux distribution conforms to the log-normal distribution have a strong linear RMS-Flux relation. However, the reverse is not true, as the log-normal distribution of flux is not a necessary result of a linear RMS-Flux relation. In the model of “minijets in a jet” indicates that the sample RMS is proportional to the sample flux if and only if the flux is the exponential of an underlying variable, as is the case for a log-normally distributed flux. This RMS-Flux relationship is a necessary and insufficient condition for the log-normal distribution.

\normalem
\begin{acknowledgements}
This work is funded by the National Natural Science Foundation of China (grants 12063007, 11863007).
This research acknowledges the Fermi Gamma-ray Space Telescope and utilizes archive data obtained from it.

\end{acknowledgements}

\bibliographystyle{raa}
\bibliography{bibtex}

\end{document}